\documentclass[letterpaper, english, reprint, aps, prb, 
	citeautoscript, groupedaddress, superscriptaddress] {revtex4-1}

\usepackage[T1]{fontenc}
\usepackage[latin9]{inputenc}
\usepackage{babel}
\usepackage{textcomp}
\usepackage[Symbol]{upgreek}
\usepackage{calc}
\usepackage{amssymb}
\usepackage{amsmath}
\usepackage{graphicx}
\usepackage{adjustbox}
\usepackage{xfrac}
\usepackage{bm}
\usepackage{dsfont}

\usepackage[colorlinks=true,linkcolor=black, citecolor=blue, urlcolor=blue, unicode=true]{hyperref}


\begin{document}
\preprint{}

\title{Field dependence of non-reciprocal magnons in chiral MnSi}

\newcommand{\tum}{Physik-Department, Technische Universit\"at M\"unchen (TUM), James-Franck-Str. 1, 85748 Garching, Germany}
\newcommand{\mlz}{Heinz-Maier-Leibnitz-Zentrum (MLZ), Technische Universit\"at M\"unchen (TUM), Lichtenbergstr. 1, 85747 Garching, Germany}
\newcommand{\lns}{Laboratory for Neutron Scattering and Imaging, Paul Scherrer Institut (PSI), CH-5232 Villigen, Switzerland}
\newcommand{\qm}{Laboratory for Quantum Magnetism, \'Ecole Polytechnique F\'ed\'erale de Lausanne, CH-1015 Lausanne, Switzerland}
\newcommand{\juelichill}{J\"ulich Centre for Neutron Science (JCNS), Forschungszentrum J\"ulich GmbH, Outstation at Institut Laue-Langevin, Bo\^ite Postale 156, 38042 Grenoble Cedex 9, France}
\newcommand{\cologne}{Institut f\"ur Theoretische Physik, Universit\"at zu K\"oln, Z\"ulpicher Str. 77a, 50937 K\"oln, Germany}
\newcommand{\dresden}{Institut f\"{u}r Theoretische Physik, Technische Universit\"{a}t Dresden, D-01062 Dresden, Germany}
\newcommand{\ill}{Institut Laue-Langevin (ILL), 71 avenue des Martyrs, 38000 Grenoble, France}

\author{T. Weber}
\email[Corresponding author: ]{tobias.weber@tum.de}
\altaffiliation[Now at: ]{\ill}
\affiliation{\tum}
\affiliation{\mlz}

\author{J. Waizner}
\affiliation{\cologne}

\author{G. S. Tucker}
\affiliation{\lns}
\affiliation{\qm}

\author{R. Georgii}
\affiliation{\mlz}
\affiliation{\tum}

\author{M. Kugler}
\affiliation{\tum}
\affiliation{\mlz}

\author{A. Bauer}
\affiliation{\tum}

\author{C. Pfleiderer}
\affiliation{\tum}

\author{M. Garst}
\affiliation{\dresden}

\author{P. B\"oni}
\affiliation{\tum}

\date{\today}

\begin{abstract}
Spin waves in chiral magnetic materials are strongly influenced by the Dzyaloshinskii-Moriya interaction resulting in intriguing phenomena like non-reciprocal magnon propagation and magnetochiral dichroism. Here, we study the non-reciprocal magnon spectrum of the archetypical chiral magnet MnSi and its evolution as a function of magnetic field covering the field-polarized and conical helix phase. Using inelastic neutron scattering, the magnon energies and their spectral weights are determined quantitatively after deconvolution with the instrumental resolution.
In the field-polarized phase the imaginary part of the dynamical susceptibility $\chi''(\varepsilon, {\bf q})$ is shown to be asymmetric with respect to wavevectors ${\bf q}$ longitudinal to the applied magnetic field ${\bf H}$, which is a hallmark of chiral magnetism. In the helimagnetic phase, $\chi''(\varepsilon, {\bf q})$ becomes increasingly symmetric with decreasing ${\bf H}$ due to the formation of helimagnon bands and the activation of additional spinflip and non-spinflip scattering channels. The neutron spectra are in excellent quantitative agreement with the low-energy theory of cubic chiral magnets with a single fitting parameter being the damping rate of spin waves.

\vspace{0.2cm}
\noindent This is a pre-print of our paper at \url{https://link.aps.org/doi/10.1103/PhysRevB.97.224403}, \\
\copyright{} 2018 American Physical Society.
\end{abstract}

\maketitle

\section{Introduction}

The dispersion of spin waves in conventional magnetic materials with an inversion center are symmetric with respect to the wavevector ${\bf q}$, $\varepsilon({\bf q}) = \varepsilon(-{\bf q})$. Such an inversion center is however absent in chiral magnets
which implies in general an asymmetric spin wave dispersion $\varepsilon({\bf q}) \neq \varepsilon(-{\bf q})$ \cite{Melcher:1973,Kataoka:1987}.
As a consequence, magnons with wavevectors ${\bf q}$ and $- {\bf q}$ possess different group velocities giving rise to non-reciprocal magnon propagation, which has been experimentally demonstrated e.g.~using spin wave spectroscopy on LiFe$_5$O$_8$ \cite{Iguchi:2015}, Cu$_2$OSeO$_3$ \cite{Seki:2016}, FeGe and Co-Zn-Mn alloys \cite{Takagi2017}. Similarly, in an inelastic scattering experiment where a certain wavevector ${\bf q}$ is transferred, magnons might be emitted with energy $\varepsilon({\bf q})$ but cannot be absorbed at the same energy, which has been observed in the chiral magnet MnSi\cite{Shirane83,Sato16} as well as in the chiral antiferromagnet $\alpha$-Cu$_2$V$_2$O$_7$ \cite{2017arXiv170204889G}. In the presence of a magnetoelectric coupling, the damping of electromagnetic waves by non-reciprocal magnons also leads to magnetochiral dichroism, for example, in Cu$_2$OSeO$_3$ for microwave frequencies\cite{Okamura:2013,Mochizuki:2015,Okamura:2015,MochizukiSeki:2015}.
Here, we demonstrate the field-dependent development of non-reciprocal spin waves by combining novel high-resolution neutron spectroscopy measurements with a well-established theoretical framework.

A material class of particular interest comprises the cubic chiral magnets with crystal symmetry P2$_1$3 including
MnSi, Cu$_2$OSeO$_3$, FeGe, and Fe$_{1-x}$Co$_x$Si. The advantage of this class of systems for investigating the spin dynamics is the relatively high symmetry of the cubic crystalline environment that considerably restricts the form of the low-energy theory in the limit of weak spin-orbit coupling. As a result, practically parameter-free predictions for the magnon spectrum are available\cite{Garst2017}. The Dzyaloshinskii-Moriya interaction (DMI) in these systems not only gives rise to a non-reciprocal magnon spectrum but also leads to spatially modulated magnetic ground states, i.e., a helix and a skyrmion lattice\cite{BauerPfleiderer2016}. The Bragg scattering of magnons from these periodic magnetic textures results in a magnon band structure where the reciprocal lattice vectors of the associated Brillouin zone is determined by the DMI.

At the $\Gamma$-point of this Brillouin zone various resonances arise due to the backfolding of the spectrum but only few of them are magnetically active.
In the helimagnetic phase, there are two magnetic resonances with finite spectral weight where the mean magnetization precesses clockwise or counterclockwise \cite{1977:Date:JPhysSocJpn,Kataoka:1987}. The skyrmion lattice phase is characterized by three magnetic resonances including a breathing mode where the mean magnetization possesses an oscillating component along the applied magnetic field \cite{2012:Mochizuki:PhysRevLett,2012:Onose:PhysRevLett}.
Quantitative agreement between theory and experiment was obtained for the  microwave resonances observed in MnSi, Fe$_{1-x}$Co$_x$Si, and Cu$_2$OSeO$_3$ \cite{Schwarze15,Weiler2017}.

The low-energy magnon band structure of the helimagnetic phase for finite wavevectors ${\bf q}$ was studied in MnSi at zero field with inelastic neutron scattering by Janoschek {\it et al.} \cite{Jano10}.
However, the obtained spectra were complex superpositions of magnon dispersions associated with the presence of multiple domains. The band structure could be resolved by Kugler \textit{et al.} \cite{Max15} after preparing a single helimagnetic domain in MnSi using a small polarizing magnetic field,
and the energies of the observed helimagnons were in quantitative agreement with theory. MnSi is especially well suited for a study of the helimagnon band structure because its characteristic energy scale is on the order of $0.1$ meV that can be resolved with state-of-the-art neutron spectrometers. For Cu$_2$OSeO$_3$, in contrast, the typical helimagnon band widths are an order of magnitude smaller limiting inelastic neutron scattering studies to high-energy features of the magnon spectrum \cite{Portnichenko:2016ey}.
In the field-polarized phase, early studies of the magnon dispersion\cite{Ishikawa77} of MnSi were not sensitive enough to observe the salient feature of non-reciprocity. It was however established later by Shirane {\it et al.} \cite{Shirane83} and recently confirmed by Grigoriev {\it et al.}\cite{Grigoriev15} and Sato {\it et al.}\cite{Sato16}.
Moreover the non-reciprocity of magnetic excitations was also demonstrated in the paramagnetic phase of MnSi using polarized neutrons by Roessli {\it et al.}\cite{Roessli2002}.

In the present work, we study the evolution of the magnon spectrum in MnSi as a function of magnetic field covering the helimagnetic phase and the field-polarized phase. Our results illustrate the development of the magnetic structure factor across the continuous phase transition at the critical field $H_{c2}$ separating the two phases elucidating symmetry aspects of the magnon dispersion and its non-reciprocity. Remarkably, taking the resolution of the neutron spectrometers into account the full neutron spectra are quantitatively explained by the theory for cubic chiral magnets using a single fitting parameter $\Gamma(H)$ for each field $H$ characterizing the intrinsic linewidth of the magnons.

In the following section \ref{sec:methods} we first describe the instruments and the experimental conditions.
In section \ref{sec:theory} we introduce the theoretical framework and provide an overview over the theoretically expected neutron spectra and their weights. In section \ref{sec:expresults} the experimental data are presented for various configurations and magnetic fields, and we conclude in section \ref{sec:disc} with a discussion.

\section{ \label{sec:methods} Experimental Methods }

Our experiments were conducted using the cold-neutron triple-axis spectrometers MIRA \cite{Robert15, miranew} at the Maier-Leibnitz-Zentrum (MLZ) in Garching, Germany, and TASP \cite{Semadeni01} at the Paul-Scherrer-Institut in Villigen, Switzerland (PSI), employing a cylindrical MnSi single-crystal ($r = 5\,\mathrm{mm}$, $h = 30\,\mathrm{mm}$) oriented with the $[001]$ direction along the cylinder axis. All inelastic scans were performed in the vicinity of the $(110)$ Bragg reflection. The energy of the incident ($E_i$) and the scattered ($E_f$) neutrons at MIRA and TASP was fixed at $E_i = 4.06$ meV and $3.5\,{\rm meV} \le E_f \le 4.06 \,{\rm meV}$, respectively. At both instruments, a neutron guide defined the divergence of the neutrons impinging the monochromator. The collimation before and after the sample was 30' and 40' for MIRA and TASP, respectively, yielding an energy resolution in the range $37\,\mu{\rm eV} \le \triangle E \le 42\,\mu{\rm eV}$. Higher order neutrons were removed by a cooled Be filter.

Before each series of measurements at fixed magnetic field $\bf H$, the sample was heated above the critical temperature into the paramagnetic phase to remove any history dependent effects. After cooling down the sample to the measurement temperature $T = 20$ K, the magnetic field $\bf H$ was applied.

For the data analysis, the neutron spectra were deconvoluted using the open-source\footnote{T. Weber, \textit{Takin} software package, code available online: \url{https://github.com/t-weber/takin}.} software package \textit{Takin}\cite{Takin2016, Takin2017, PhDWeber}, which employs the algorithm by Eckold and Sobolev \cite{Eckold2014} to calculate the four-dimensional instrumental resolution function. Using a four-dimensional deconvolution procedure we gain quantitative information from our experimental data involving not only the magnon energies, but also their spectral weights. This is especially important in the vicinity of the phase transition at $H_{c2}$ where the helimagnetic bands would otherwise be indistinguishable due to their close proximity to each other in energy.

\section{Theory of neutron scattering cross section in chiral magnets}
\label{sec:theory}

\subsection{Low-energy theory of cubic chiral magnets}

The cubic chiral magnets at low energies are described  by the free energy density\cite{Max15,Garst2017} $\mathcal{F} = \mathcal{F}_0 + \mathcal{F}_{\rm dipolar} + \mathcal{F}_{\rm corr}$ that depends on the magnetization ${\bf M} = m \hat n$ with the amplitude $m$ and the unit vector $\hat n$. The exchange contribution at low energies is given by\cite{Bak1980}
\begin{align} \label{Theory}
\mathcal{F}_0 = \frac{\rho_s}{2} \left[ (\nabla_i \hat n_j)^2 - 2 k_{h0} \hat n (\nabla \times \hat n)\right] - \mu_0 m \hat n {\bf H}
\end{align}
with the exchange stiffness density $\rho_s$ and the applied field ${\bf H}$. The contribution $\mathcal{F}_{\rm dipolar}$ is due to dipolar interactions and $\mathcal{F}_{\rm corr}$ is associated with higher-energy corrections. In Ref.~\onlinecite{Max15} the latter term was chosen to be of the form $\mathcal{F}_{\rm corr} = \frac{\rho_s}{2} \frac{\mathcal{A}}{k_h^2} (\nabla^2 \hat n)^2$. At high fields $H > H_{c2}$, the ground state is field-polarized $\hat n = {\bf H}/|{\bf H}|$, and at lower fields $H < H_{c2}$ a conical helix emerges.

For a field applied along the $z$-axis, the helix is given by $\hat n(z) = (\sin \theta \cos(k_h z), -\sin \theta \sin(k_h z), \cos \theta)$, and it is left-handed for the sign of the Dzyaloshinskii-Moriya interaction ($k_{h0} > 0$) chosen in Eq.~\eqref{Theory}.
In zeroth order in the correction $\mathcal{A}$, the pitch vector $k_h$ is determined by $k_h = k_{h0}$;
a finite $\mathcal{A}$ slightly renormalizes the value of $k_h$.
The cone angle $\cos \theta = H/H_{c2}$ is determined by the magnetic field interpolating smoothly between conical and field-polarized phase. The precessional dynamics of the magnetization is governed by the equation of motion
$\partial_t \hat n = - \frac{g\mu_B}{\hbar} \hat n \times {\bf B}_{\rm eff}$ with the effective field ${\bf B}_{\rm eff} = -\frac{1}{m} \frac{\delta F}{\delta \hat n}$ with $F = \int d{\bf r} \mathcal{F}$.
The stiffness of the magnon dispersion is given by $\mathcal{D} = g\mu_B \rho_s/m = g\mu_B \mu_0 H^{\rm int}_{c2}/k_h^2$, that can be expressed in terms of the internal critical field
$H^{\rm int}_{c2} = H_{c2} - N m$ with the appropriate demagnetization factor $N$.
Our experiments were performed at a temperature $T = 20$ K on MnSi where the pitch vector $k_h = 0.036$ \AA$^{-1}$, the internal critical field $\mu_0 H^{\rm int}_{c2} = 0.53$ T, the susceptibility $\chi^{\rm int}_{\rm con} = m/H^{\rm int}_{c2} = 0.34$ and the stiffness $\mathcal{D} = 47.8$ meV \AA$^2$ with the $g$-factor $g \approx 2$ for MnSi \cite{Max15, Jano10}. Moreover, the high-energy correction was determined in Ref.~\onlinecite{Max15} to be $\mathcal{A} = - 0.0073$.\footnote{Strictly speaking, the negative value of $\mathcal{A}$ requires further higher-order corrections in order to stabilize the theory.} This parameter set completely fixes the magnon dispersion providing a parameter-free prediction as a function of applied field ${\bf H}$.

\begin{figure*}
\centering
\includegraphics[width=1.7\columnwidth]{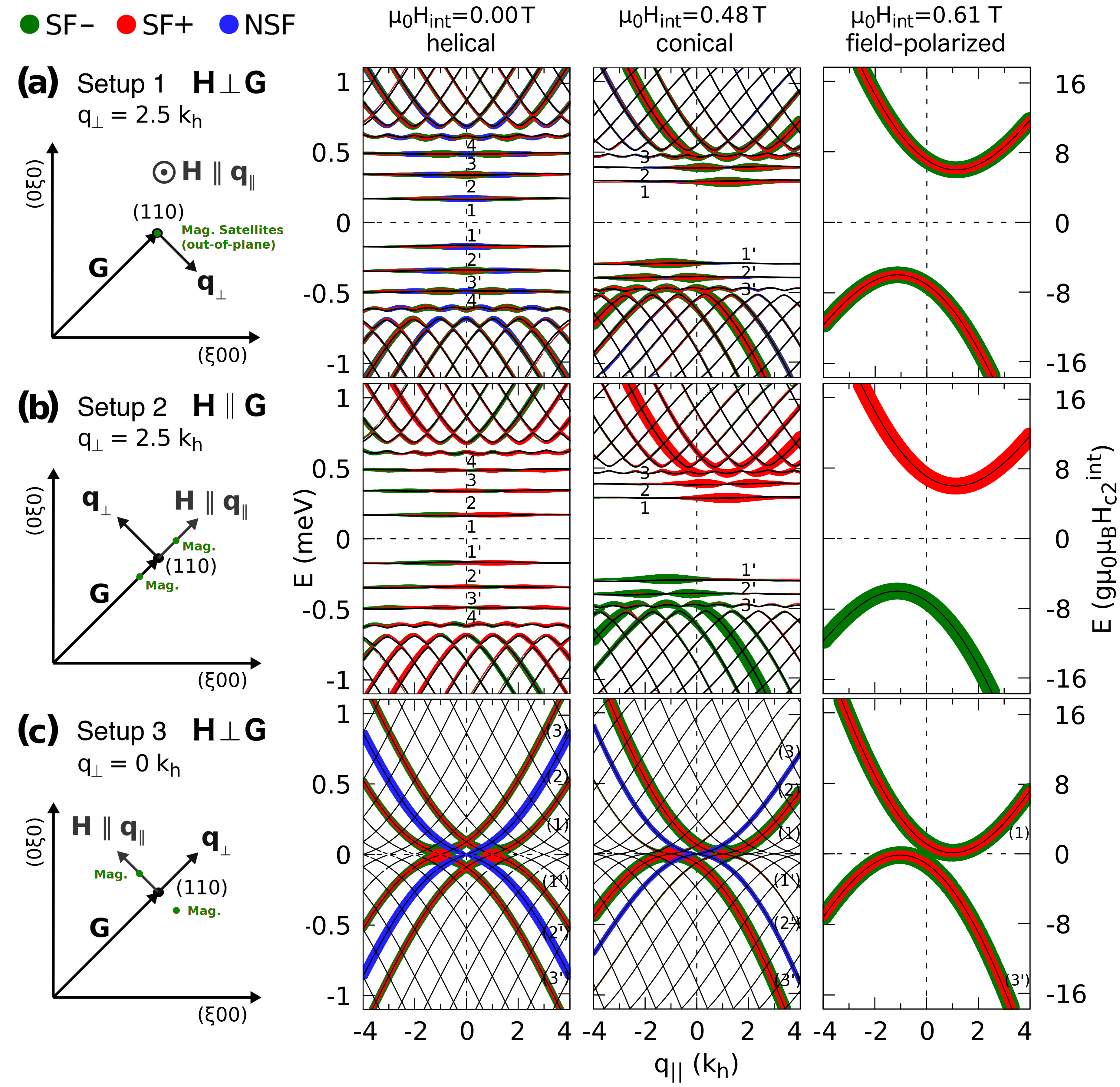}
\caption{\label{fig1}
{\bf Evolution of the magnon spectrum and spectral weights in a magnetic field ${\bf H}$.}
The three spectra for each setup represent the helical phase at zero field, the conical phase at intermediate fields, and the field-polarized phase at $\mu_0 H_{\rm int} > \mu_0 H^{\rm int}_{c2} = 0.53$ T. The magnon wavevector $q_\parallel$ parallel to the field is shown on the horizontal axis while ${\bf q}_\perp \perp {\bf H}$ is fixed.
Panels {\bf (a)} and {\bf (b)} show the spectrum for a fixed $|{\bf q}_\perp| = 2.5 k_h$ and for an applied field perpendicular and along the nuclear reciprocal lattice vector ${\bf G}$, respectively. Panel {\bf (c)} shows the spectrum for a fixed ${\bf q}_\perp = 0$ and  ${\bf H}\perp {\bf G}$. The numbers indicate the observed peaks in Figs.~\ref{figSetup1} - \ref{fig4}. The theoretically expected spectral weights are represented by the shaded red and green lines corresponding to spinflip (SF$\pm$) scattering and blue lines corresponding to non-spinflip (NSF) scattering  processes. The thickness of the shading scales linearly with the spectral weight except in panel {\bf (c)} in close proximity to the magnetic satellites $q_\parallel = \pm k_h$ where the spectral weight diverges and the shading instead is scaled logarithmically with the weight. Note that there is a finite magnon gap in the field-polarized phase that is, however, still small for the chosen field value $\mu_0 H_{\rm int} = 0.61$ T.}
\end{figure*}

The differential cross section for inelastic neutron scattering is given by
\begin{align} \label{dsigma}
\frac{d^2 \sigma}{d \varepsilon d\Omega} \propto (1+n_B(\varepsilon))\, {\rm tr}\{(\mathds{1} - \hat G \hat G^T)
\chi''(\varepsilon,{\bf q})\}
\end{align}
where ${\bf q}$ is the wavevector transfer with respect to a nuclear reciprocal lattice vector ${\bf G}$ with $|{\bf q}| \ll |{\bf G}|$. The Bose factor $n_B$ ensures that the absorption of magnons with energy $\varepsilon$ is suppressed at low temperatures. The projection operator $P_{\hat G} = \mathds{1} - \hat G \hat G^T$ arises from dipolar interactions
between the neutron spin and the magnetization projecting onto the space perpendicular to the unit vector $\hat G = {\bf G}/|{\bf G}|$. All our experiments were performed with respect to $\hat G = (1,1,0)/\sqrt{2}$.
We obtain the imaginary part of the susceptibility matrix $\chi''_{ij}$ of the magnetization with the help of linear spinwave theory \cite{Belitz06, Petrova11, Max15}. In principle, the trace in Eq.~\eqref{dsigma} can be decomposed into a sum of three contributions consisting of two spinflip processes and one non-spinflip process. For a neutron-spin polarized along the magnetic field axis $\hat H = {\bf H}/|{\bf H}|$ the component $\hat H^T P_{\hat G} \chi'' P_{\hat G} \hat H$ describes the non-spinflip scattering event and does not contribute for $\bf H \parallel \bf G$ as $P_{\hat G} \hat H = 0$.  In the following discussion, it is instructive to distinguish these processes although in our unpolarized scattering experiment all of them are added up according to Eq.~\eqref{dsigma}.

\subsection{Magnon spectrum and spectral weights}
The theoretically expected magnon spectrum and the associated spectral weights are illustrated in Fig.~\ref{fig1} for the three experimental setups used. In setup 1 and 3 the magnetic field, ${\bf H} \perp {\bf G}$, is applied perpendicular to the nuclear reciprocal lattice vector and ${\bf H} \parallel {\bf G}$ in setup 2. The dispersion is shown as a function of the magnon wavevector $q_\parallel$ parallel to the applied field for a fixed wavevector ${\bf q}_\perp$ perpendicular to ${\bf H}$ that is $|{\bf q}_\perp| = 2.5 k_h$ for setup 1 and 2 and ${\bf q}_\perp = 0$ for setup 3. The temperature independent spectral weights from the trace in Eq.~\eqref{dsigma} associated with two spinflip scattering processes and a single non-spinflip process are represented by the red, green and blue shades, respectively. Whereas the spectrum of setup 1 and 2 is the same, the weight distribution differs because the non-spinflip scattering does not contribute in setup 2 as ${\bf H} \parallel {\bf G}$. The energy transfer is shown on the vertical axis where positive and negative energies, respectively, correspond to the creation and absorption of a magnon by the neutron.

Three representative spectra are shown in Fig.~\ref{fig1} for each setup: for the helical phase at zero field, the conical phase at finite field $H_{\rm int} < H^{\rm int}_{c2}$, and the field-polarized phase for $H_{\rm int} > H^{\rm int}_{c2}$. Below the critical field $H^{\rm int}_{c2}$, the periodic magnetic texture leads to a band structure for the magnons.
However, due to the screw symmetry of the helix magnon band gaps only appear for a finite perpendicular wavevector ${\bf q}_\perp$ as otherwise the Bragg scattering off the periodic magnetization associated with helical ordering is inactive\cite{Garst2017}. For $|{\bf q}_\perp| = 2.5 k_h$ the lowest three bands are basically flat at zero field and, in addition, the weights of the three channels are distributed over various helimagnon bands.

As the magnetic field increases the weight of the non-spinflip scattering process in setup 1 decreases and vanishes at the critical field $H^{\rm int}_{c2}$. In the field-polarized phase only a single magnon branch remains that is only sensitive to spinflip scattering in our setups. This branch corresponds to a parabola at low energies that is shifted by the DMI giving rise to a non-reciprocal spectrum. Whereas in setup 1 both spinflip scattering channels contribute equally, in setup 2 the weight of this branch is dominated by a single spinflip process.
For vanishing ${\bf q}_\perp = 0$ in setup 3 the band gaps are absent  at zero field and the scattering weight is limited to three distinct helimagnon branches, which are centered around the nuclear Bragg peak ($q_\parallel = 0$) and the two magnetic satellite peaks ($q_\parallel = \pm k_h$) of the static structure factor.

With increasing field  the weight of the non-spinflip contribution again decreases and vanishes in the field-polarized phase where only a single branch survives. Similar to setup 1, both spinflip processes contribute equal weight to this branch at $H_{\rm int} > H^{\rm int}_{c2}$. Moreover, the magnon excitations acquire a finite gap at $q_\parallel = \pm k_h$ when the field exceeds $H^{\rm int}_{c2}$.

The distribution of spectral weight  reflects the symmetry of the dynamic susceptibility $\chi_{ij}''(\varepsilon,{\bf q}) = -\chi_{ji}''(-\varepsilon,-{\bf q})$. Accordingly, the emission of a magnon with energy $\varepsilon({\bf q})$, for example, in one spinflip channel (red shaded) possesses the same weight as the absorption of a magnon with energy $\varepsilon(-{\bf q})$ in the other spinflip channel (green shaded).
In the field-polarized phase $H > H_{c2}$ the spectrum exhibits a pronounced non-reciprocity, $\varepsilon({\bf q}_\perp,q_\parallel) \neq \varepsilon({\bf q}_\perp, - q_\parallel)$, but only with respect to the wavevector $q_\parallel$ longitudinal to the field. Neglecting magnetocrystalline anisotropies, there exists a combined rotation symmetry in spin and real space around the magnetic field axis ensuring that the magnon energy, $\varepsilon(|{\bf q}_\perp|, q_\parallel)$, only depends on the amplitude but not on the direction of ${\bf q}_\perp$.

In the helimagnetically ordered phase, this rotation symmetry is broken by the magnetic ground state. However, the helix still possesses the screw symmetry so that the magnon energy $\varepsilon(|{\bf q}_\perp|, q_\parallel)$ still depends only on the amplitude of ${\bf q}_\perp$ even for $H_{\rm int} < H^{\rm int}_{c2}$. The non-reciprocity $\varepsilon(|{\bf q}_\perp|, q_\parallel) \neq \varepsilon(|{\bf q}_\perp|, -q_\parallel)$
is still present at finite field and materializes, for example in panel {\bf (c)}, as a shift of the band crossings in the conical phase away from $q_\parallel = 0$. This shift can be attributed to the higher-energy correction $\mathcal{F}_{\rm corr}$ and thus becomes less important at low frequencies.

Finally, at zero magnetic field the helix $\hat n(z) = (\cos(k_h z), -\sin(k_h z), 0)$ is also invariant with respect to a $\pi$-rotation of spin and real space around the $x$-axis which ensures that $\varepsilon(|{\bf q}_\perp|, q_\parallel) = \varepsilon(|{\bf q}_\perp|, -q_\parallel)$ so that the spectrum becomes reciprocal at ${\bf H} = 0$. Whereas the spectrum is reciprocal at zero field, the weight distribution might remain non-reciprocal. For example, in a {\it polarized} neutron scattering experiment with a polarization longitudinal to $\bf G$ the weight distribution for spinflip scattering processes will be asymmetric in $q_\parallel$ even for ${\bf H} = 0$ (cf.~setup 2).

\section{Experimental results}
\label{sec:expresults}

\subsection{Helimagnon bands at finite wavevector ${\bf q}_\perp \perp {\bf H}$}

\begin{figure*}
\includegraphics[width=2\columnwidth]{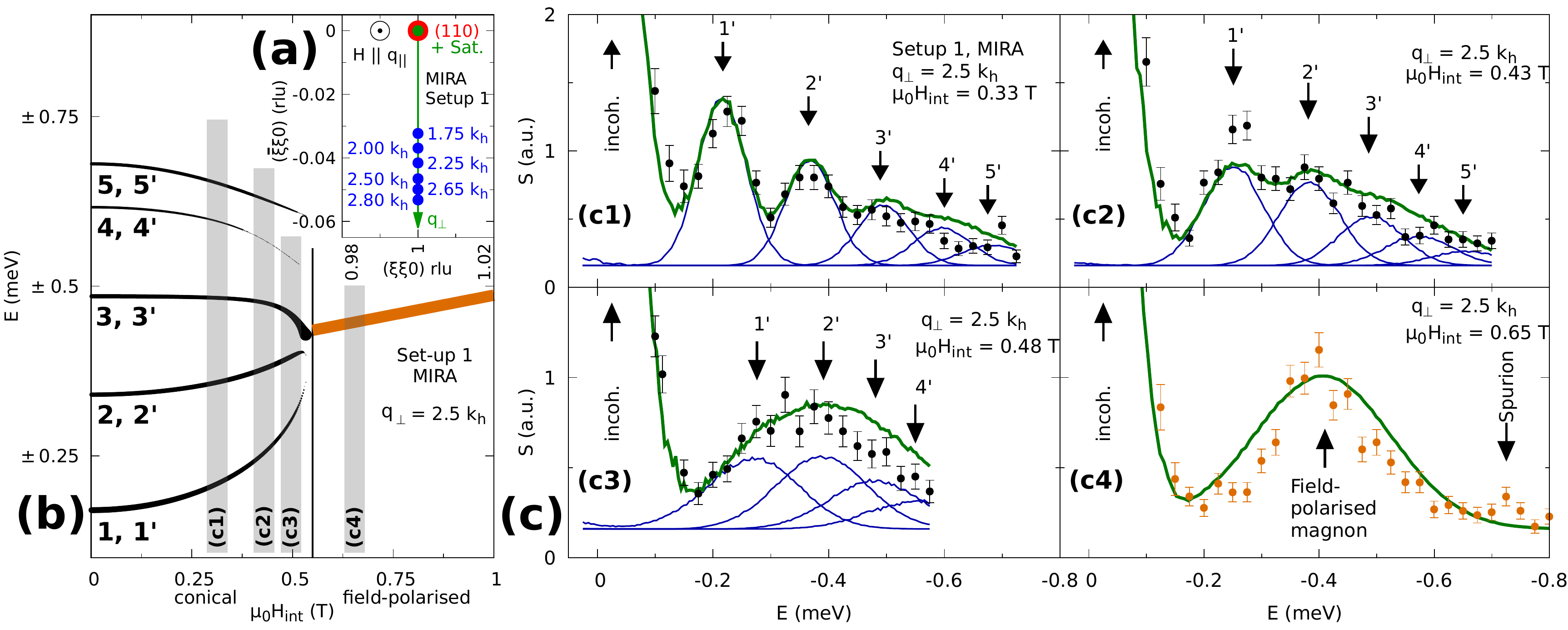}
\caption{{\bf Helimagnon bands at finite ${\bf q}_\perp$ with ${\bf H} \perp {\bf G}$.}
Panel {\bf (a)}: Data was collected at different transferred momenta ${\bf q}_\perp$ (blue dots) perpendicular to the applied magnetic field. The red dot shows the position of the nuclear Bragg reflection $\hat G = \frac{1}{\sqrt{2}}(110)$ with ${\bf H} \perp {\bf G}$. The green dot depicts the projections of the out-of-plane helimagnetic satellite peaks. Panel {\bf (b)}: Field-dependence of the theoretically expected total weight at $|{\bf q}_\perp| = 2.5 k_h$
corresponding to a cut through the spectra of Fig.~\ref{fig1} {\bf a} at $q_\parallel = 0$. For unpolarized neutrons the weight at $q_\parallel = 0$ is the same for positive and negative energy transfer.
 Panel {\bf (c)}: The experimental data (dots) for different magnetic fields and a comparison to theory (green lines) after convolution with the experimental resolution. The blue lines and arrows indicate the position of the individual helimagnon bands taking into account a finite linewidth $\Gamma$. The peak numbers indicate the numbering of the dispersion branches in Fig.~\ref{fig1} {\bf a}. A spurion was identified at $\mu_0 H_{\rm int} = 650$ mT. \cite{Supplement} }
\label{figSetup1}
\end{figure*}

\begin{figure*}
\includegraphics[width=2\columnwidth]{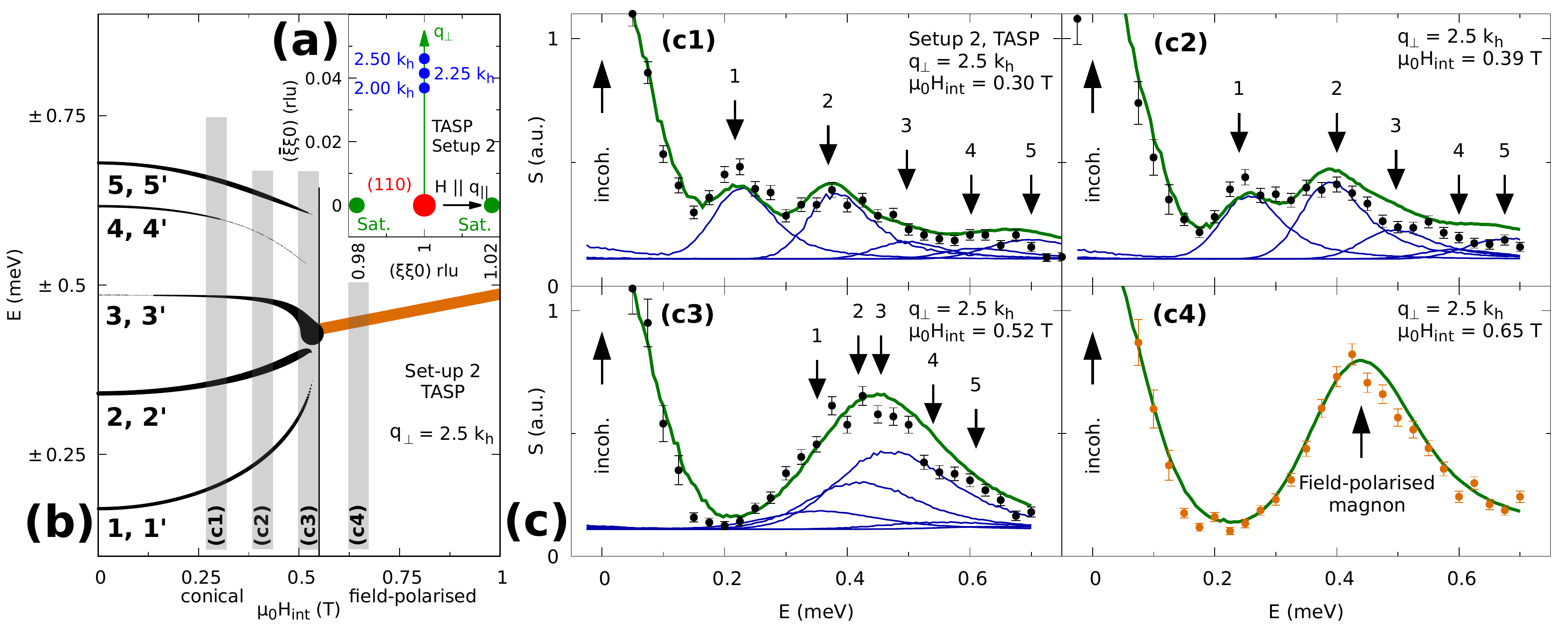}
\caption{{\bf Helimagnon bands at finite ${\bf q}_\perp$ with ${\bf H} \parallel {\bf G}$.}
The content of the panels are similar to Fig.~\ref{figSetup1} but here ${\bf H} \parallel {\bf G}$, leading to a different distribution of spectral weights. The green dots in panel {\bf (a)} indicate positions of the magnetic satellite peaks due to helimagnetic order. The peak numbers indicate the numbering of the dispersion branches in Fig.~\ref{fig1} {\bf b}. }
\label{figSetup2}
\end{figure*}

In order to probe the magnon spectrum at a finite wavevector ${\bf q}_\perp$ perpendicular to the applied field, experiments at the instruments MIRA and TASP (see section \ref{sec:methods}) were performed with a magnetic field along the crystallographic $[001]$ and $[110]$-direction corresponding to setup 1 with ${\bf H} \perp {\bf G}$ and setup 2 with ${\bf H} \parallel {\bf G}$, respectively. A finite field was applied in order to prepare a state with a single magnetic domain only. Measurements were taken at different values of ${\bf q}_\perp$, see panel {\bf (a)} of Figs.~\ref{figSetup1} and \ref{figSetup2}. Here we concentrate on the results with the reduced momentum transfer $|{\bf q}_\perp| = 2.5 k_h$ and refer to the supplement \cite{Supplement} for the other values. The longitudinal momentum was chosen to be zero so that the obtained neutron data correspond to cuts through the spectra of Fig.~\ref{fig1} {\bf a} and {\bf b} at $q_\parallel = 0$.

\begin{figure*}
\includegraphics[width=2\columnwidth]{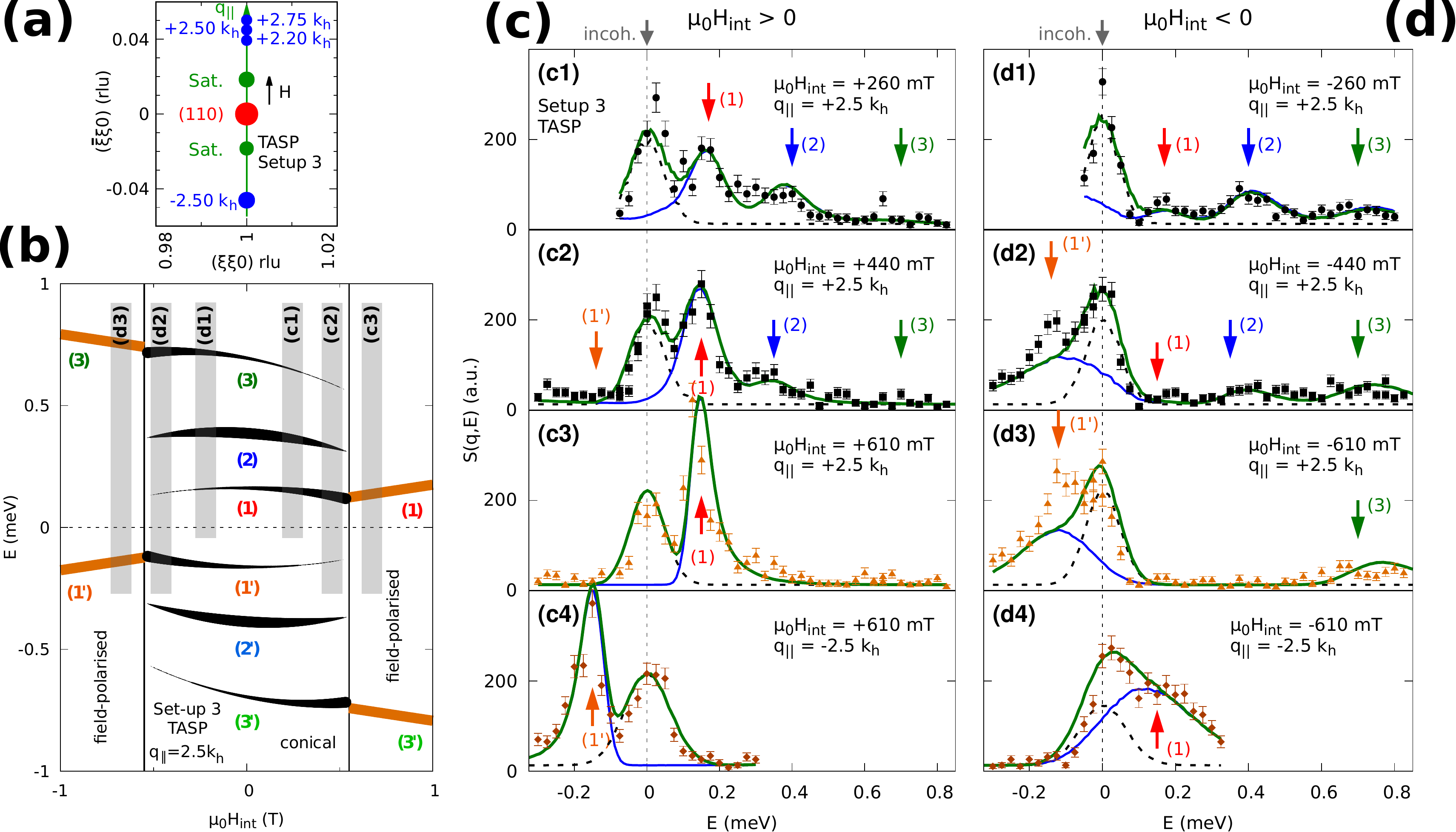}
\caption{{\bf Non-reciprocal helimagnon dispersion at finite $q_\parallel$.}
 Panel {\bf (a)}: Data was collected at different transferred momenta $q_\parallel$ (blue dots) longitudinal to the applied magnetic field. The red dot shows the position of the nuclear Bragg reflection $\hat G = \frac{1}{\sqrt{2}}(110)$ with ${\bf H} \perp {\bf G}$. The green dots indicate the positions of the magnetic Bragg peaks of helimagnetic order. Panel {\bf (b)}: Field-dependence of the theoretically expected weight at ${\bf q}_\perp = 0$ and $q_\parallel = 2.5 k_h$ corresponding to cuts through the spectra of Fig.~\ref{fig1} {\bf c}. For unpolarized neutrons, the weight at zero field is the same for positive and negative  energy transfer whereas non-reciprocity develops for finite ${\bf H}$.
Panel {\bf (c)}: The experimental data (dots) for different positive magnetic fields and a comparison with theory (green lines) after convolution with the experimental resolution. Three branches are detected in the helimagnetically ordered phase whereas a single branch is present in the field-polarized phase for $\mu_0 H_{\rm int}> \mu_0 H^{\rm int}_{c2} = 530$ mT (arrows). Panel {\bf (d)}: Experimental data for different negative magnetic fields. Comparison with panel {\bf (c)} demonstrates the non-reciprocity of the spectrum under reversal of ${\bf H}$. The non-reciprocity with respect to $q_\parallel$ is also explicitly demonstrated for $\mu_0 H_{\rm int} = \pm 610$ mT in sub-panels {\bf (c4)} and {\bf (d4)}. The peak numbers indicate the numbering of the dispersion branches in Fig.~\ref{fig1} {\bf c}. }
\label{fig4}
\label{figSetup3}
\end{figure*}

The evolution of the spectra with field as measured in setup 1 is shown in Fig.~\ref{figSetup1}. Panel {\bf (b)} shows the theoretically expected weight as a function of field and panel {\bf (c)} displays the neutron spectra. The data was collected at negative energies via absorption of magnons.
The green lines were obtained by a convolution of the theoretical spectrum with the resolution function of the instrument.
The blue lines indicate the contributions of individual helimagnon bands whose positions are also indicated by arrows.
In addition, an intrinsic linewidth $\Gamma$ has been introduced which is approximated to be the same for all bands. The linewidth $\Gamma$ is the single fitting parameter as all other parameters were taken from previous measurements.
At low magnetic fields four helimagnon bands are resolved. With increasing field, these bands come closer in energy and start to merge as the critical field $H_{c2}$ is approached. Above the critical field, a single magnon excitation survives in the field-polarized state.

Fig.~\ref{figSetup2} shows the corresponding spectra for setup 2 which were collected at positive energy transfer through emission of magnons. Due to the parallel alignment ${\bf H} \parallel {\bf G}$ the non-spinflip scattering does not contribute. As a result, the weight of the third band at low fields is suppressed in comparison to setup 1. The third band gains importance as the field increases and substantially contributes to the total weight close to the critical field. Above $H_{c2}$ again only a single magnon mode is detected.

\subsection{Non-reciprocal helimagnon dispersion for $q_\parallel \parallel {\bf H}$}

The magnon spectrum for vanishing ${\bf q}_\perp$ but finite wavevector $q_\parallel$ along the field was experimentally probed with setup 3 at the TASP spectrometer with ${\bf H}$ applied along $[\bar 110]$ so that ${\bf H} \perp {\bf G}$.
This configuration corresponds to cuts through the spectra of Fig.~\ref{fig1} {\bf c} at the respective values of $q_\parallel$.
As shown in Fig.~\ref{fig4} {\bf a}, different values for $q_\parallel$ were investigated \cite{Supplement} and distinguished from other spectra of non-magnetic origin \cite{Lamago10}. Here, however, we concentrate on the magnons at $q_\parallel = \pm 2.5 k_h$ only.

The measured neutron spectra are shown in panels {\bf (c)} and {\bf (d)}, and compared with the theoretically expected spectral weights, which are shown in panel {\bf (b)}. Three helimagnon branches are observed in panel {\bf (c)} below the critical field, $H < H_{c2}$. With increasing positive field, the peaks at positive energy transfer associated with branches (2) and (3) lose weight whereas branch (1) gains weight as the critical field is approached. Near $H_{c2}$, branch (1) continuously develops into the magnon excitation of the field-polarized phase. The experimental spectrum possesses a strong asymmetry with respect to the sign of the energy transfer (see peaks (1) and (1')) that is intrinsic and cannot be accounted for by the Bose factor of Eq.~\eqref{dsigma}. This is explicitly demonstrated by measurements with reversed transferred momentum $q_\parallel$ as shown in sub-panels {\bf c4} and {\bf d4}. Instead of peak (1) at positive energy transfer, a strong magnon peak (1') appears at negative energy transfer confirming the non-reciprocity of the spectrum.

The non-reciprocity of the dynamic structure factor becomes also apparent after reversing the magnetic field.
The reversal of ${\bf H}$ effectively inverts the magnon spectrum $\varepsilon(q_\parallel) \to \varepsilon(-q_\parallel)$ and also reshuffles the spectral weight. In panel {\bf (d)} the neutron spectra are shown for negative fields that are to be compared with the results of panel {\bf (c)}. Non-reciprocity is only expected at finite fields. However, for $\mu_0 H_{\rm int} = -260$ mT the spectrum is already clearly distinct from the one at positive field. The branch (2) has the largest spectral weight at ${\bf H} = 0$. With increasing field its weight, however, decreases and the branch (3) gains in importance. It is this latter branch that now continuously connects with the magnon mode of the field-polarized phase. The degree of non-reciprocity of the neutron spectrum increases with increasing $|{\bf H}|$ and is particularly pronounced for $|{\bf H}|>H_{c2}$.

\section{ \label{sec:disc} Discussion }

\begin{figure}
\includegraphics[width=0.975\columnwidth]{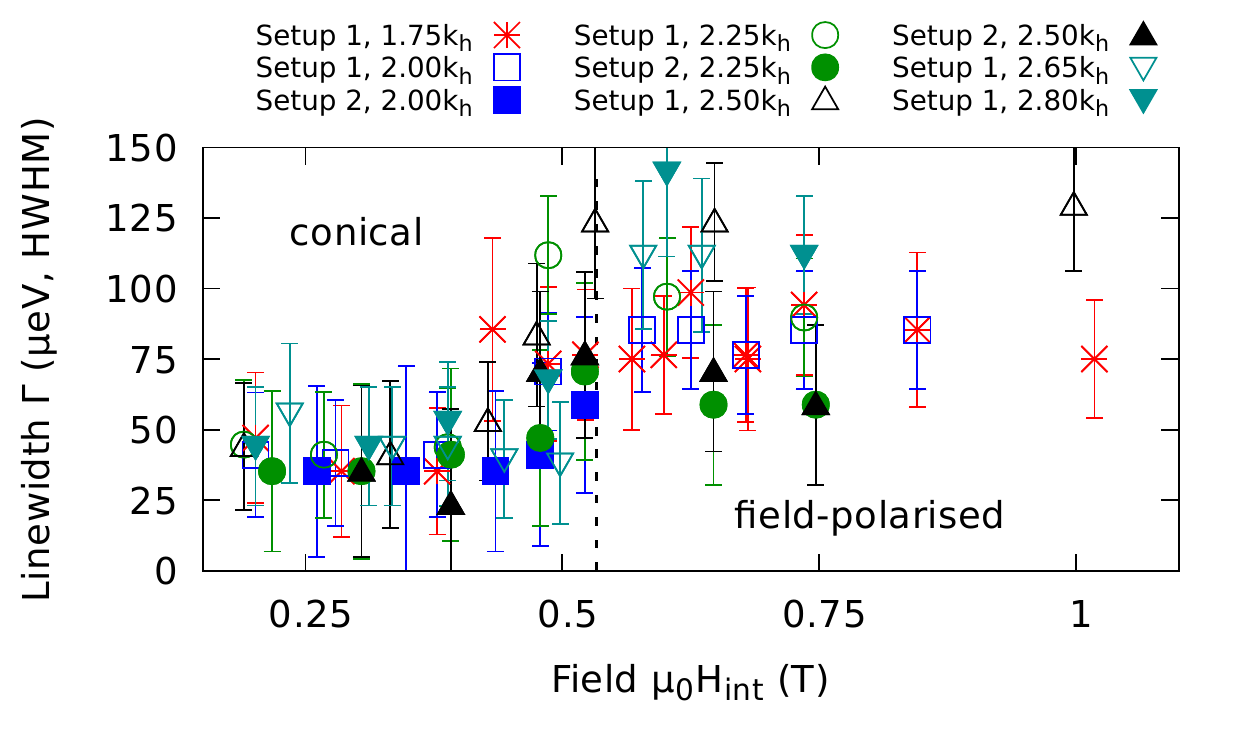}
\caption{\label{fig:linewidths} {\bf Magnon linewidth $\Gamma$.} The values of the half-width at half-maximum (HWHM)
were obtained from fitting the spectra of
setups 1 and 2 at different transferred momenta ${\bf q}_\perp$ and $q_\parallel = 0$.}
\end{figure}

We investigated the magnon spectrum of MnSi for various values of the magnetic field covering the helimagnetically ordered as well as the field-polarized phase using inelastic neutron scattering.
For all investigated values of the magnetic field we find excellent quantitative agreement between the experimental neutron scattering data and theory after convolution with the instrumental resolution function\cite{Takin2016, Takin2017, PhDWeber}; the resulting quantitative theoretical spectra are shown as green lines in Figs.~\ref{figSetup1} {\bf c}, \ref{figSetup2} {\bf c} and \ref{figSetup3} {\bf c,d}. Remarkably, both the dispersion $\varepsilon({\bf q})$ of the magnon resonances as well as their spectral weights agree with parameter-free theoretical predictions. All parameters of the low-energy theory, see section \ref{sec:theory}, were determined by previous measurements.

After taking into account the full four-dimensional instrumental resolution function, the quantitative comparison between theory and experiment allowed us to extract also a field-dependent linewidth $\Gamma(H)$ (HWHM) of the resonances, which is the single fitting parameter necessary for the quantitative description of the experimental structure factor. A single value $\Gamma(H)$ was sufficient to describe the data of all magnon resonances for a given setup sufficiently well. Interestingly, for setup 3, i.e., for vanishing ${\bf q}_\perp = 0$ the linewidth turned out to be negligible within the instrumental resolution. For setups 1 and 2, e.g., for $|{\bf q}_\perp| = 2.5 k_h$ we obtained, however, finite values for $\Gamma$ hinting at a strong dependence of the magnon lifetime on the wavevector ${\bf q}_\perp$ perpendicular to the applied magnetic field.

The finite values for $\Gamma$ fitted to the data of setups 1 and 2 are summarized in Fig.~\ref{fig:linewidths}. Values of approximately $\Gamma \sim$ 35 $\mu$eV were found in the helimagnetically ordered phase and larger values $\Gamma \sim$ 85 $\mu$eV in the field-polarized phase. For typical magnon energies of $\varepsilon \sim 0.2$ meV and $\varepsilon \sim 0.4$ meV in the helimagnetic and field-polarized phase, respectively, this corresponds to an effective damping parameter $\alpha = \Gamma/\varepsilon$ of approximately $0.17$ and $0.21$. These values for $\alpha$ are roughly a factor of three larger than the findings of Schwarze {\it et al.} \cite{Schwarze15} using magnetic resonance measurements, i.e., for spinwave excitations at ${\bf q} = 0$. This is not unexpected as the phase space for magnon decay increases with increasing wavevector. Our findings suggests that the decay rate especially increases with the wavevector ${\bf q}_\perp$ pointing in a direction perpendicular to the magnetic field. The lifetime of the spinwave modes probed in our experiments, in contrast to the uniform resonances, should be determined mostly by intrinsic effects like magnon-magnon interactions. We are not aware of any investigation of the magnon lifetime in chiral magnets, e.g., within the framework of non-linear spinwave theory
so that our results for $\Gamma$ with its strong dependence on ${\bf q}_\perp$ should be clarified in future theoretical studies.

Nevertheless, such a detailed theoretical understanding of the experimentally observed magnon dispersion and spectral weights as presented in this work is unprecedented for chiral magnets in general and, in particular, for MnSi. Previous neutron scattering studies that considered the low-energy properties of chiral spinwaves were often limited to the field-polarized phase and analyzed only the dispersion but not the spectral weight of the resonances\cite{Ishikawa77,Shirane83,Grigoriev15,Sato16}. Other works that investigated the spinwaves in the helical and conical phase were not able to resolve the helimagnon band structure \cite{Shirane83,Jano10,Portnichenko:2016ey}. Kugler \textit{et al.} \cite{Max15} succeeded to resolve single helimagnon bands at small fields and also described their dispersion theoretically  but not their spectral weights. Our results for $H \ll H_{c2}$ are fully consistent with the data of Ref.~\onlinecite{Max15}.

In conclusion, using inelastic neutron scattering we have investigated the evolution of the band structure of the helimagnons in MnSi and the emergence of non-reciprocity as a function of the magnetic field ${\bf H}$ and the sign of the momentum transfer {\bf q} covering the helical, conical helix and the field-polarized phases. The observed dispersion of the spinwaves as well as their spectral weights are quantitatively explained by the low-energy theory of chiral magnets consisting of the symmetric exchange, Dzyaloshinskii-Moriya, Zeeman, and dipolar energies. These results set the stage for future explorations of the magnetization dynamics associated with the remaining phase in the phase diagram of chiral magnets housing the topological skyrmion crystal \cite{JanoSkyrmi, fobes2018spin, SkxAsym18} as well as with the enigmatic non-Fermi liquid regime in the paramagnetic phase of MnSi at high pressures \cite{PfleidererNFL}.

\begin{acknowledgments}
This work is based upon experiments performed at the MIRA instrument operated by FRM II at the Heinz Maier-Leibnitz Zentrum (MLZ), Garching, Germany.
Furthermore, this work is based on experiments performed at the Swiss spallation neutron source SINQ, Paul Scherrer Institute, Villigen, Switzerland.
This work was part of the Ph.D. thesis of T. Weber \cite{PhDWeber} and was supported by the DFG under GE 971/5-1. M.G. is supported by the DFG via SFB 1143 ``Correlated Magnetism: From Frustration to Topology'' and grant GA 1072/5-1.
A.B. and C.P. gratefully acknowledge financial support through DFG TRR80 (project E1) and ERC Advanced Grant 291079 (TOPFIT).
We thank R. Schwikowski for technical support and G. Brandl for his implementation of the helimagnon model which he created for Ref. \onlinecite{Max15}, and which we further developed for this work.
\end{acknowledgments}


\end{document}


\preprint{}

\newcommand{\Angs}{\mathrm{\mbox{\AA}}}
\newcommand{\mnsi}{$\textrm{Mn}\textrm{Si}$}

\title{Supplemental Material -- \\Field dependence of non-reciprocal magnons in chiral MnSi}
\newcommand{\tum}{Physik-Department, Technische Universit\"at M\"unchen (TUM), James-Franck-Str. 1, 85748 Garching, Germany}
\newcommand{\mlz}{Heinz-Maier-Leibnitz-Zentrum (MLZ), Technische Universit\"at M\"unchen (TUM), Lichtenbergstr. 1, 85747 Garching, Germany}
\newcommand{\lns}{Laboratory for Neutron Scattering and Imaging, Paul Scherrer Institut, CH-5232 Villigen, Switzerland}
\newcommand{\qm}{Laboratory for Quantum Magnetism, \'Ecole Polytechnique F\'ed\'erale de Lausanne, CH-1015 Lausanne, Switzerland}
\newcommand{\juelichill}{J\"ulich Centre for Neutron Science (JCNS), Forschungszentrum J\"ulich GmbH, Outstation at Institut Laue-Langevin, Bo\^ite Postale 156, 38042 Grenoble Cedex 9, France}
\newcommand{\cologne}{Institut f\"ur Theoretische Physik, Universit\"at zu K\"oln, Z\"ulpicher Str. 77a, 50937 K\"oln, Germany}
\newcommand{\dresden}{Institut f\"{u}r Theoretische Physik, Technische Universit\"{a}t Dresden, D-01062 Dresden, Germany}
\newcommand{\ill}{Institut Laue-Langevin (ILL), 71 avenue des Martyrs, 38000 Grenoble, France}

\author{T. Weber}
\email[Corresponding author: ]{tobias.weber@tum.de}
\altaffiliation[Now at: ]{\ill}
\affiliation{\tum}
\affiliation{\mlz}

\author{J. Waizner}
\affiliation{\cologne}

\author{G. S. Tucker}
\affiliation{\lns}
\affiliation{\qm}

\author{R. Georgii}
\affiliation{\mlz}
\affiliation{\tum}

\author{M. Kugler}
\affiliation{\tum}
\affiliation{\mlz}

\author{A. Bauer}
\affiliation{\tum}

\author{C. Pfleiderer}
\affiliation{\tum}

\author{M. Garst}
\affiliation{\dresden}

\author{P. B\"oni}
\affiliation{\tum}


\date{\today}

\begin{abstract}
\noindent This is a pre-print of the supplement to our paper at \url{https://link.aps.org/doi/10.1103/PhysRevB.97.224403}, 
\copyright{} 2018 American Physical Society.
\end{abstract}

\maketitle

\section{ \label{sec:setup1} Experimental data}

We have indicated in the insets of panels {\bf (a)} of Figs. 2, 3, and 4 of the main text the positions in reciprocal space where the magnetic excitation spectra have been measured. In the following we show these measurements and compare them with the convoluted magnetic cross section of the helimagnon model.

\begin{figure*}[!htb]
    \includegraphics[trim=4bp 4bp 35bp 35bp, clip, width=0.80\textwidth]{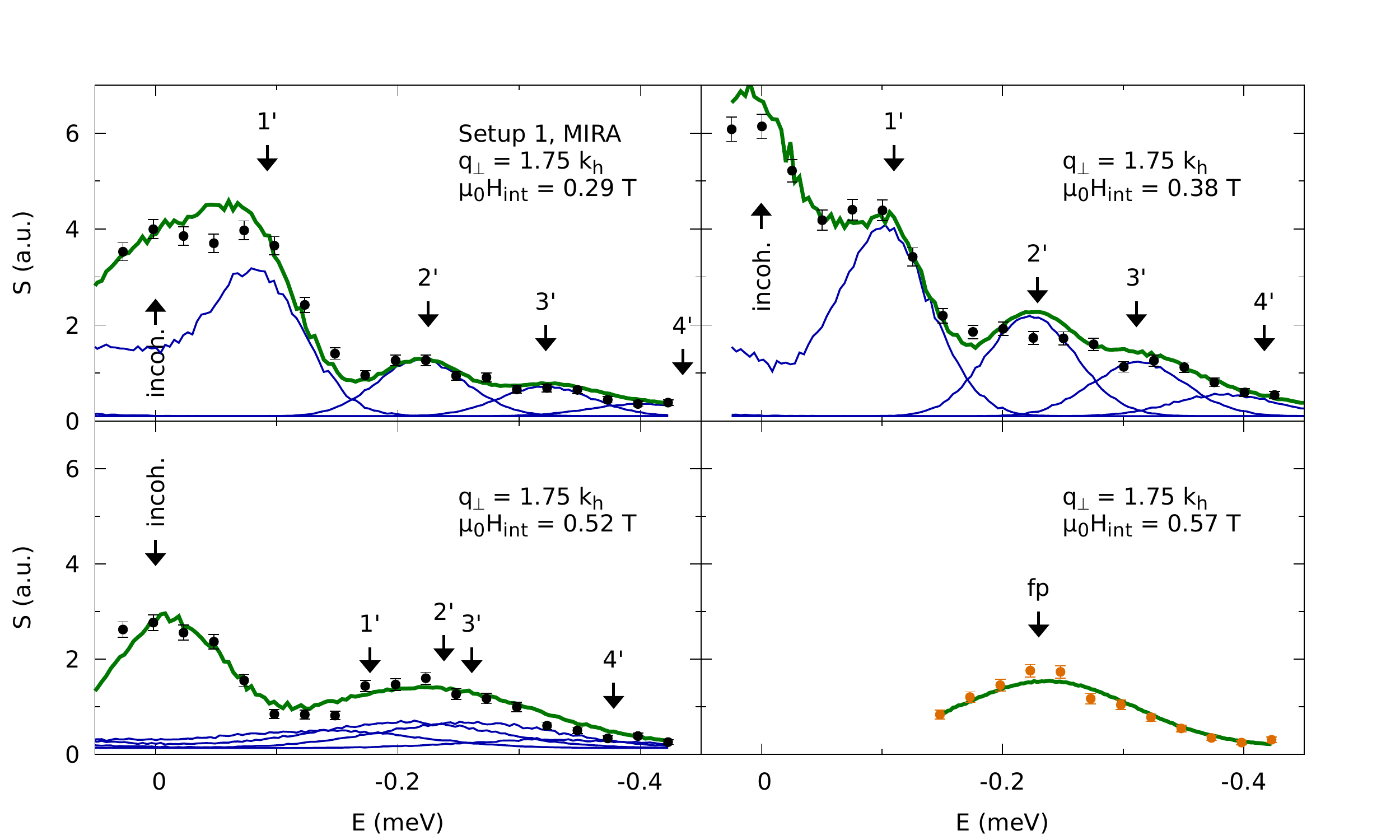}
    \caption{ Field dependence of the helimagnon bands for $|{\bf q}_\perp|=1.75\,k_h$ and $q_\parallel = 0$ as measured with ${\bf H} \perp \bf{G}$ (Setup 1).}
    \label{setup1_175}
\end{figure*}

\begin{figure*}[!htb]
    \includegraphics[trim=4bp 4bp 35bp 35bp, clip, width=0.80\textwidth]{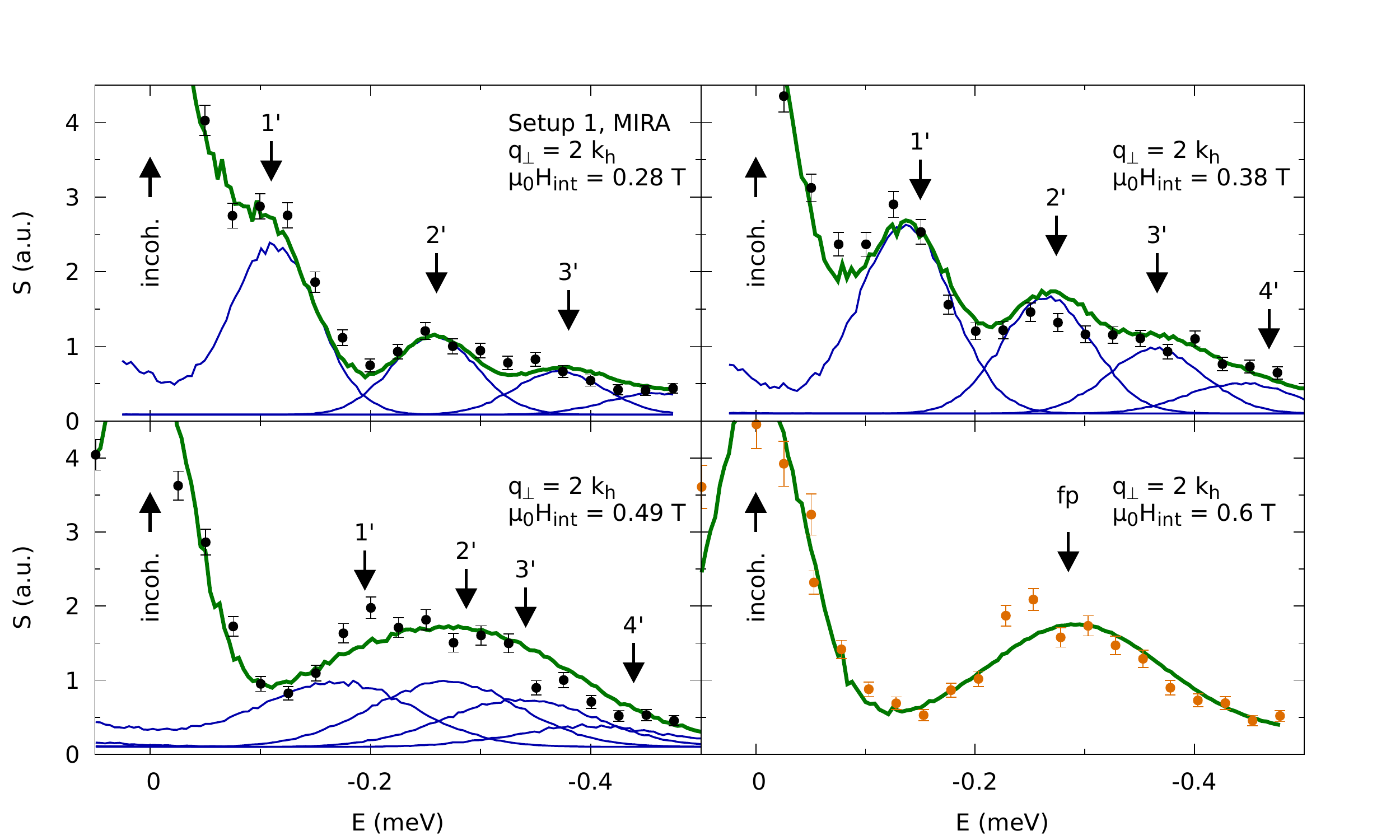}
    \caption{Field dependence of the helimagnon bands for $|{\bf q}_\perp|=2.00\,k_h$ and $q_\parallel = 0$ as measured with ${\bf H} \perp \bf{G}$ (Setup 1).}
    \label{setup1_200}
\end{figure*}

\begin{figure*}[!htb]
    \includegraphics[trim=4bp 4bp 35bp 35bp, clip, width=0.80\textwidth]{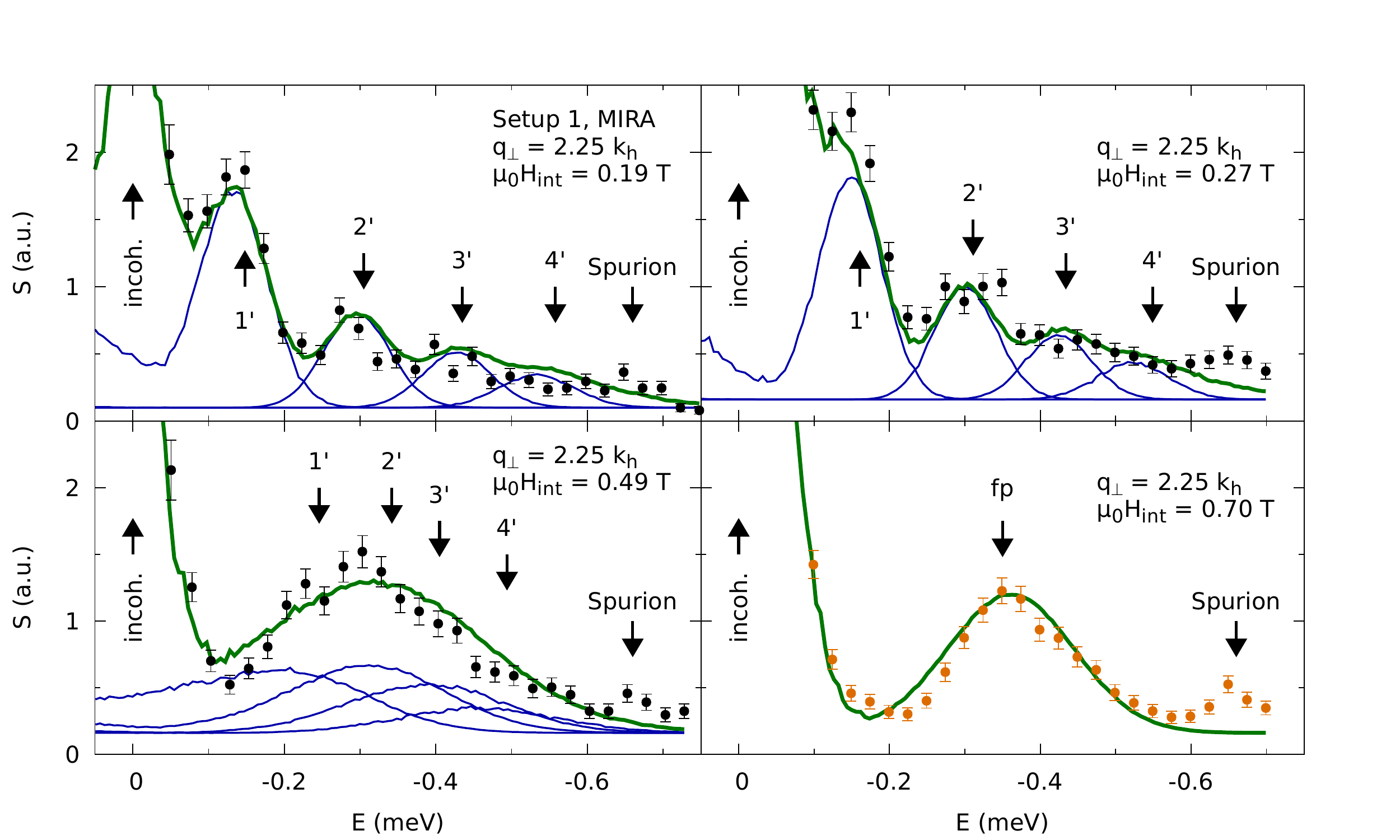}
    \caption{Field dependence of the helimagnon bands for $|{\bf q}_\perp|=2.25\,k_h$ and $q_\parallel = 0$ as measured with ${\bf H} \perp \bf{G}$ (Setup 1). A spurious, non-magnetic peak is observed at the border of the scan range (see text).}
    \label{setup1_225}
\end{figure*}

\begin{figure*}[!htb]
    \includegraphics[trim=4bp 4bp 35bp 35bp, clip, width=0.80\textwidth]{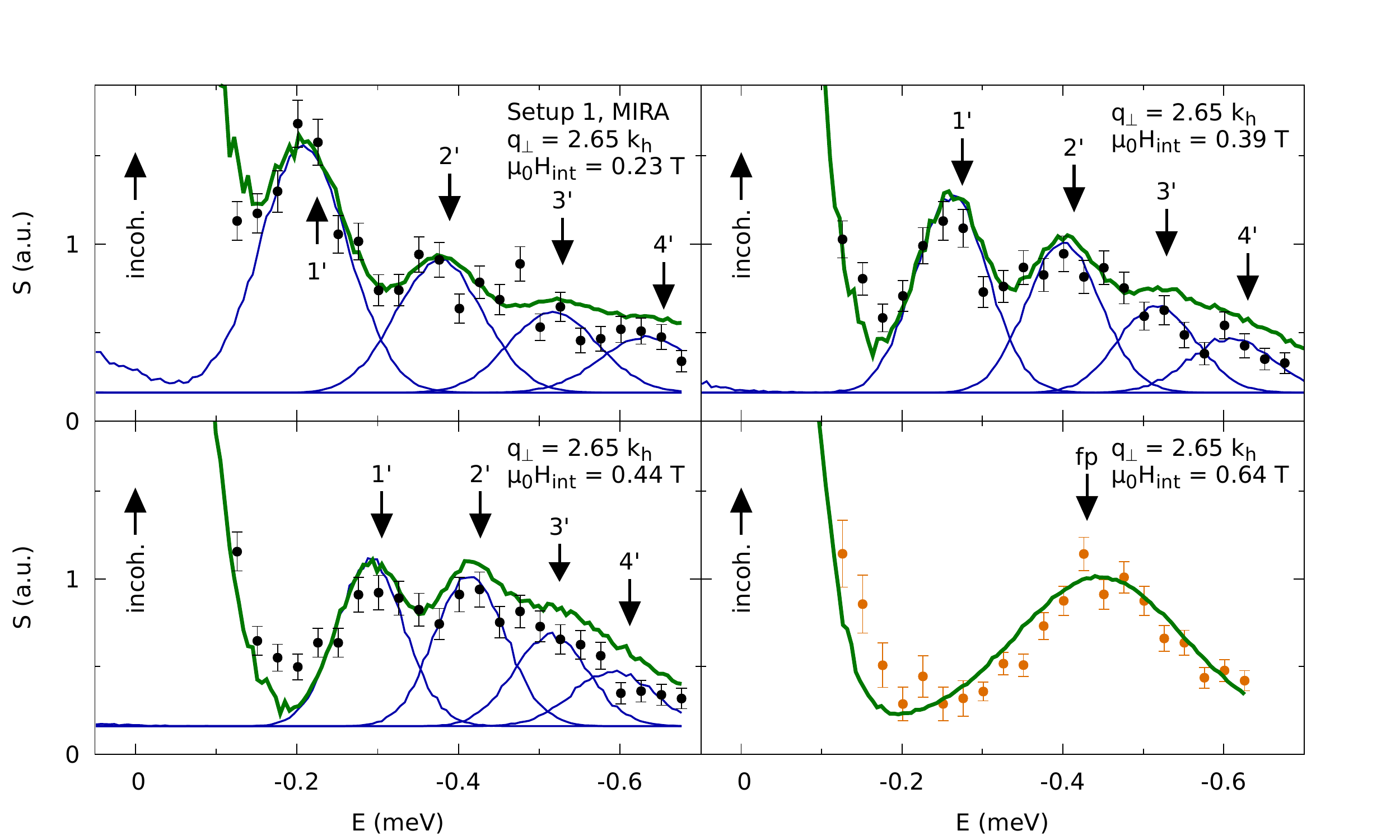}
    \caption{Field dependence of the helimagnon bands for $|{\bf q}_\perp|=2.65\,k_h$ and $q_\parallel = 0$ as measured with ${\bf H} \perp \bf{G}$ (Setup 1).}
    \label{setup1_265}
\end{figure*}

\begin{figure*}[!htb]
    \includegraphics[trim=4bp 4bp 35bp 35bp, clip, width=0.80\textwidth]{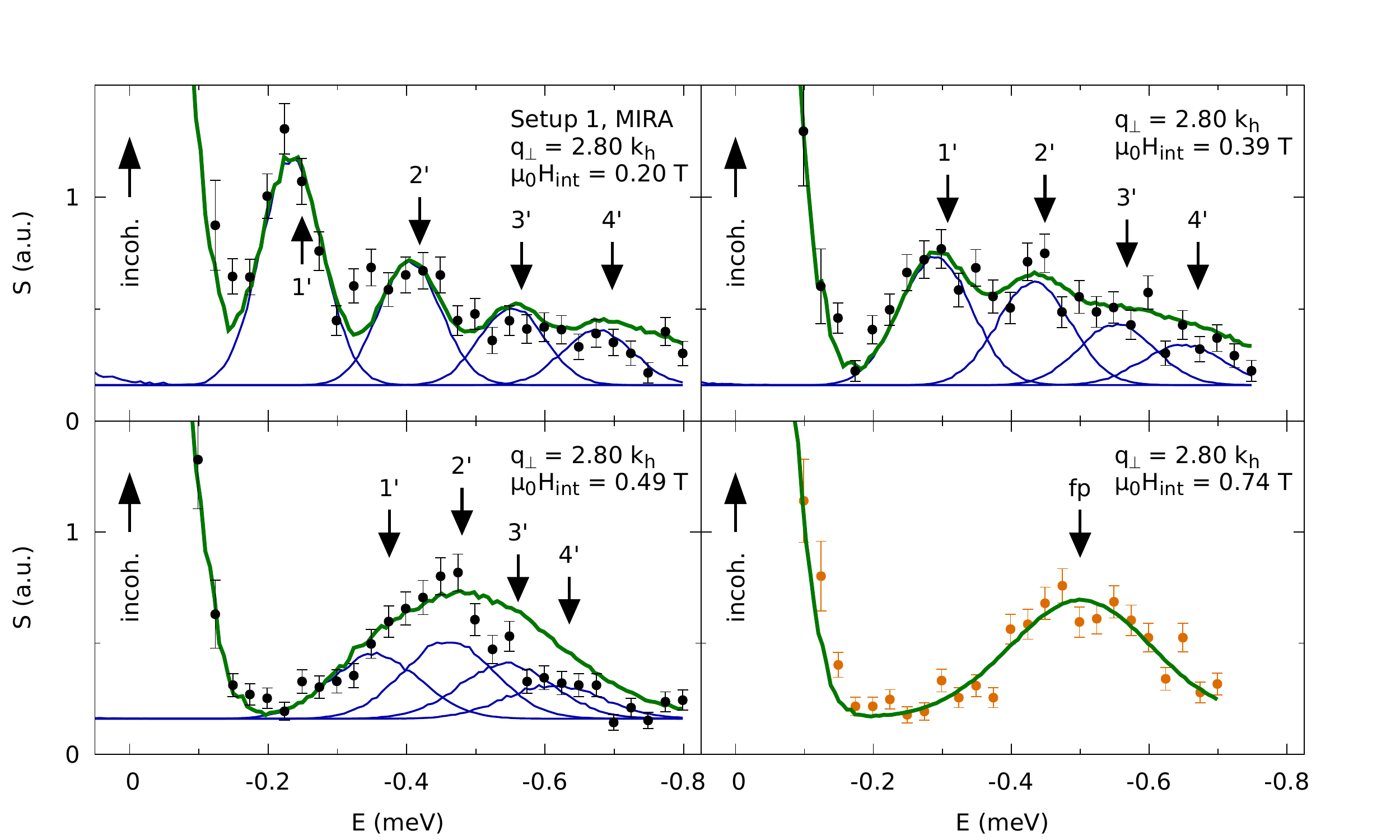}
    \caption{Field dependence of the helimagnon bands for $|{\bf q}_\perp|=2.80\,k_h$ and $q_\parallel = 0$ as measured with ${\bf H} \perp \bf{G}$ (Setup 1).}
    \label{setup1_280}
\end{figure*}

\begin{figure*}[!htb]
    \includegraphics[trim=4bp 4bp 35bp 35bp, clip, width=0.80\textwidth]{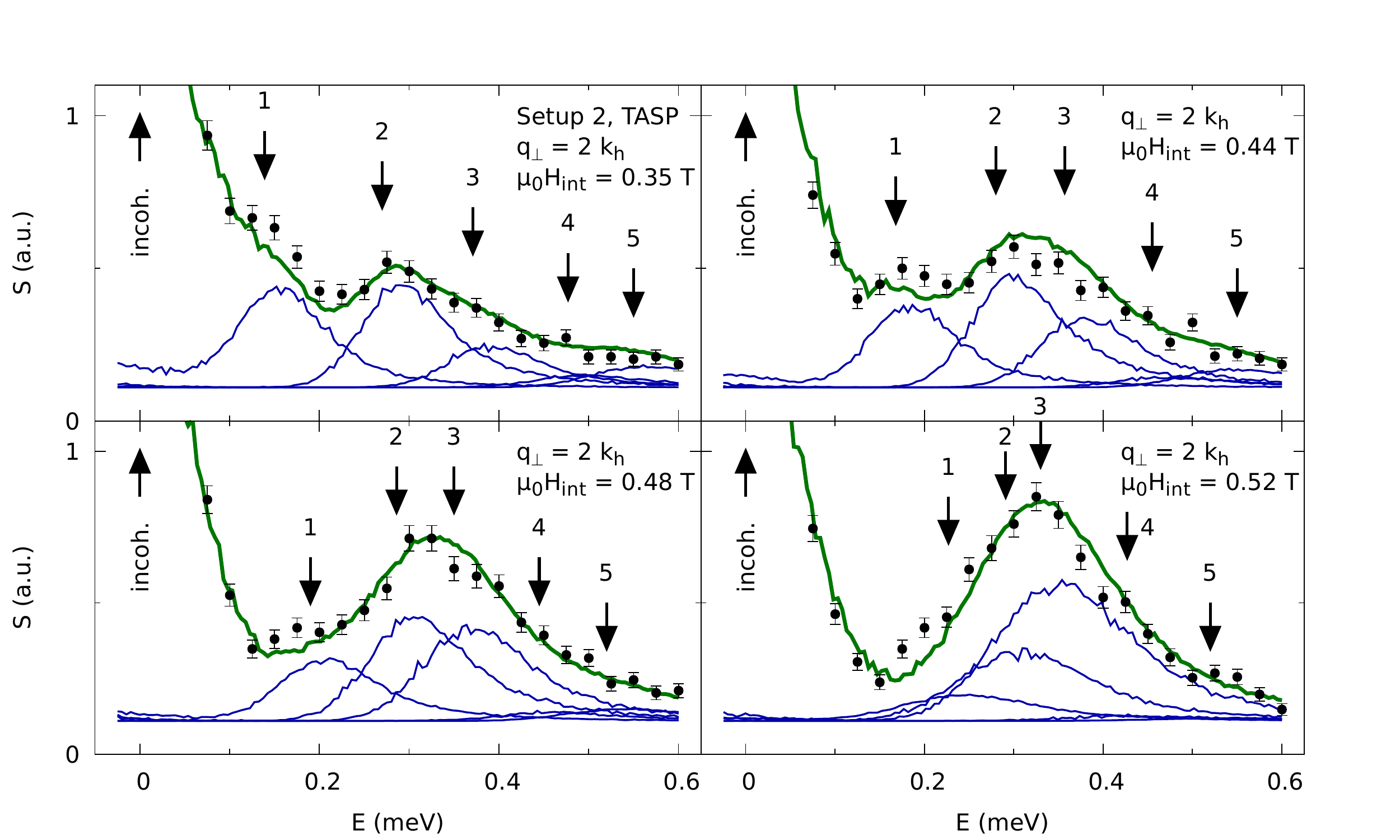}
    \caption{Field dependence of the helimagnon bands for $|{\bf q}_\perp|=2.00\,k_h$ and $q_\parallel = 0$ as measured with ${\bf H} \parallel \bf{G}$ (Setup 2).}
    \label{setup2_200}
\end{figure*}

\begin{figure*}[!htb]
    \includegraphics[trim=4bp 4bp 35bp 35bp, clip, width=0.80\textwidth]{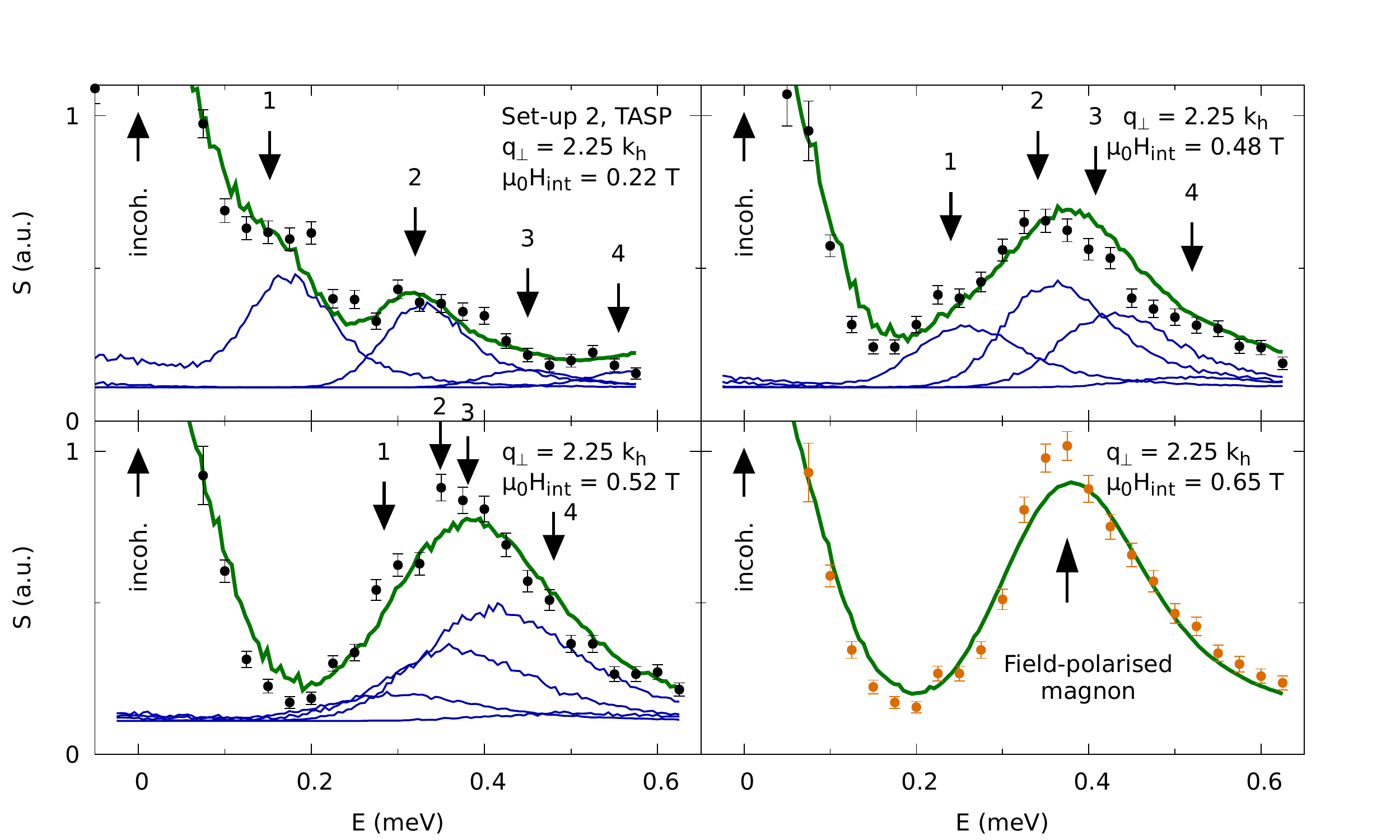}
    \caption{Field dependence of the helimagnon bands for $|{\bf q}_\perp|=2.25\,k_h$ and $q_\parallel = 0$ as measured with ${\bf H} \parallel \bf{G}$ (Setup 2).}
    \label{setup2_225}
\end{figure*}

\begin{figure*}[!htb]
    \includegraphics[width=0.9\textwidth]{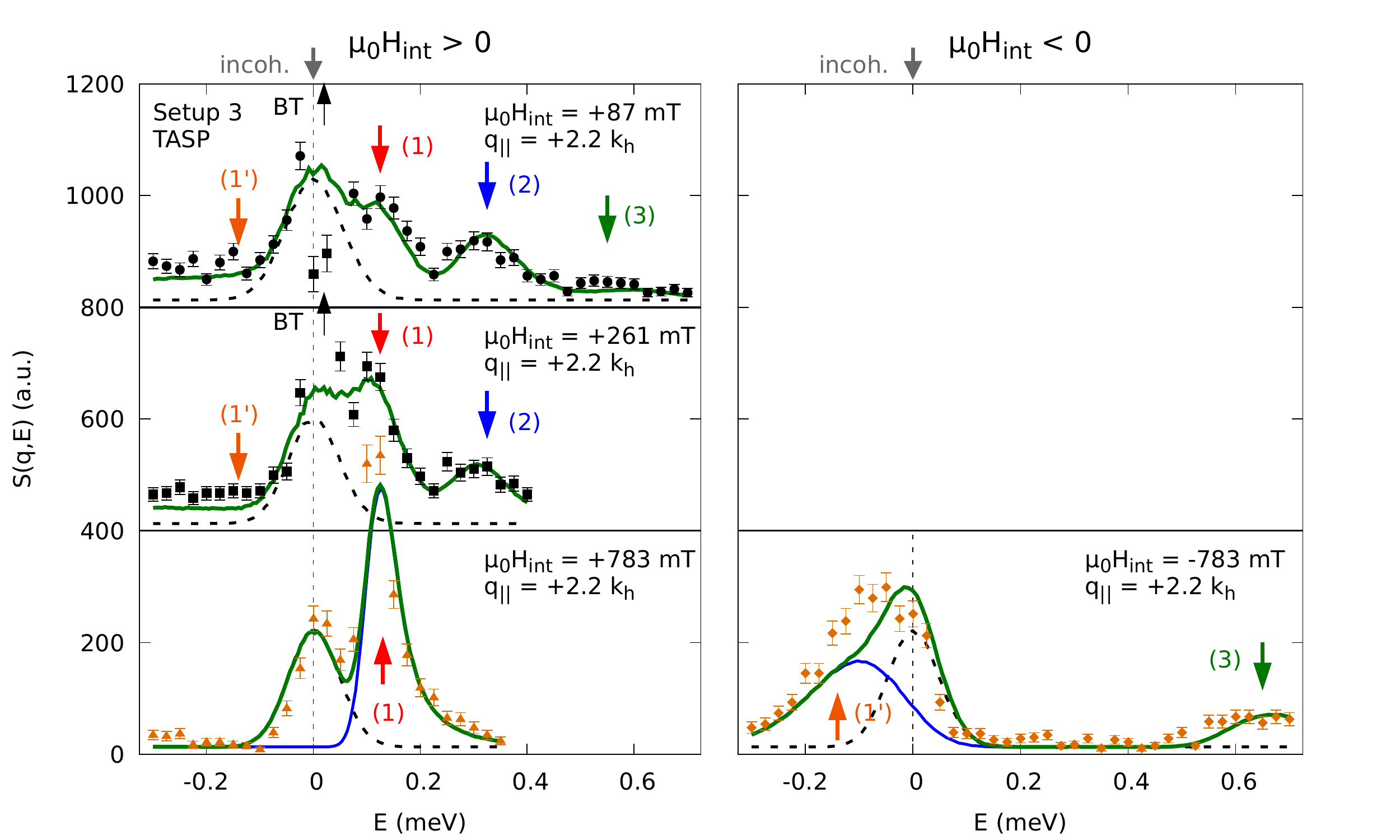}
    \caption{Field dependence of the helimagnon modes for $|{\bf q}_\parallel| = 2.2\,k_h$ and $q_\perp = 0$ as measured with ${\bf H} \perp \bf{G}$ (Setup 3). In the top-left panel ($H_{int} = 87$ mT), the intensity of three data points in the interval $0 \le E \le 0.05\,\mathrm{meV}$ is outside the depicted plot range, they constitute a spurious Bragg tail (see text), which is marked with ``BT'' and an arrow. The plots have been shifted by multiples of 400 units in the individual panels.}
    \label{setup3_220}
\end{figure*}

\begin{figure*}[!htb]
    \includegraphics[width=0.50\textwidth]{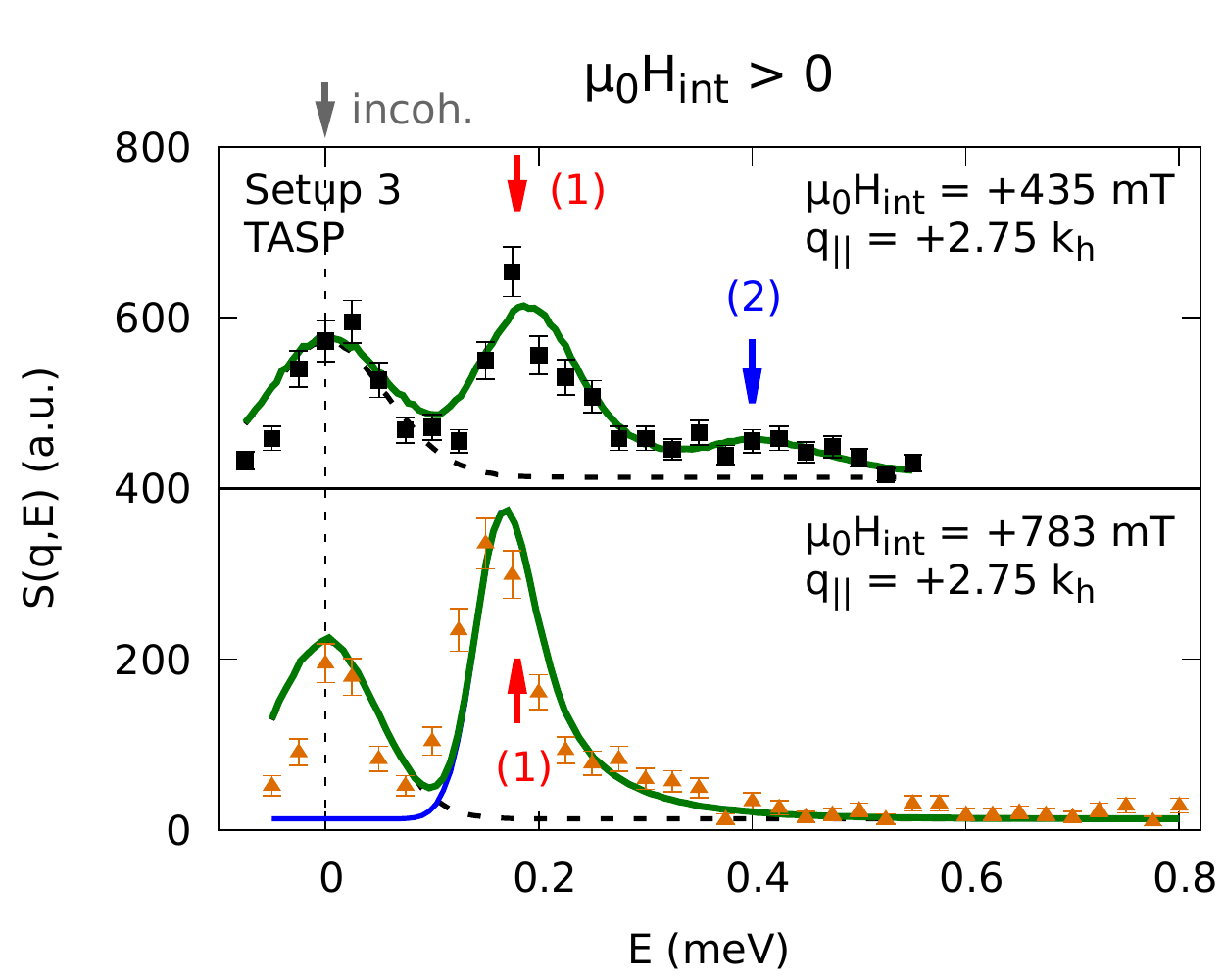}
    \caption{Field dependence of the helimagnon modes for $|{\bf q}_\parallel| = 2.75\,k_h$ and $q_\perp = 0$ as measured with ${\bf H} \perp \bf{G}$ (Setup 3). The plots have been shifted by multiples of 400 units in the individual panels.}
    \label{setup3_275}
\end{figure*}

\begin{figure*}[!htb]
    \includegraphics[trim=4bp 40bp 20bp 40bp, clip, width=0.45\textwidth]{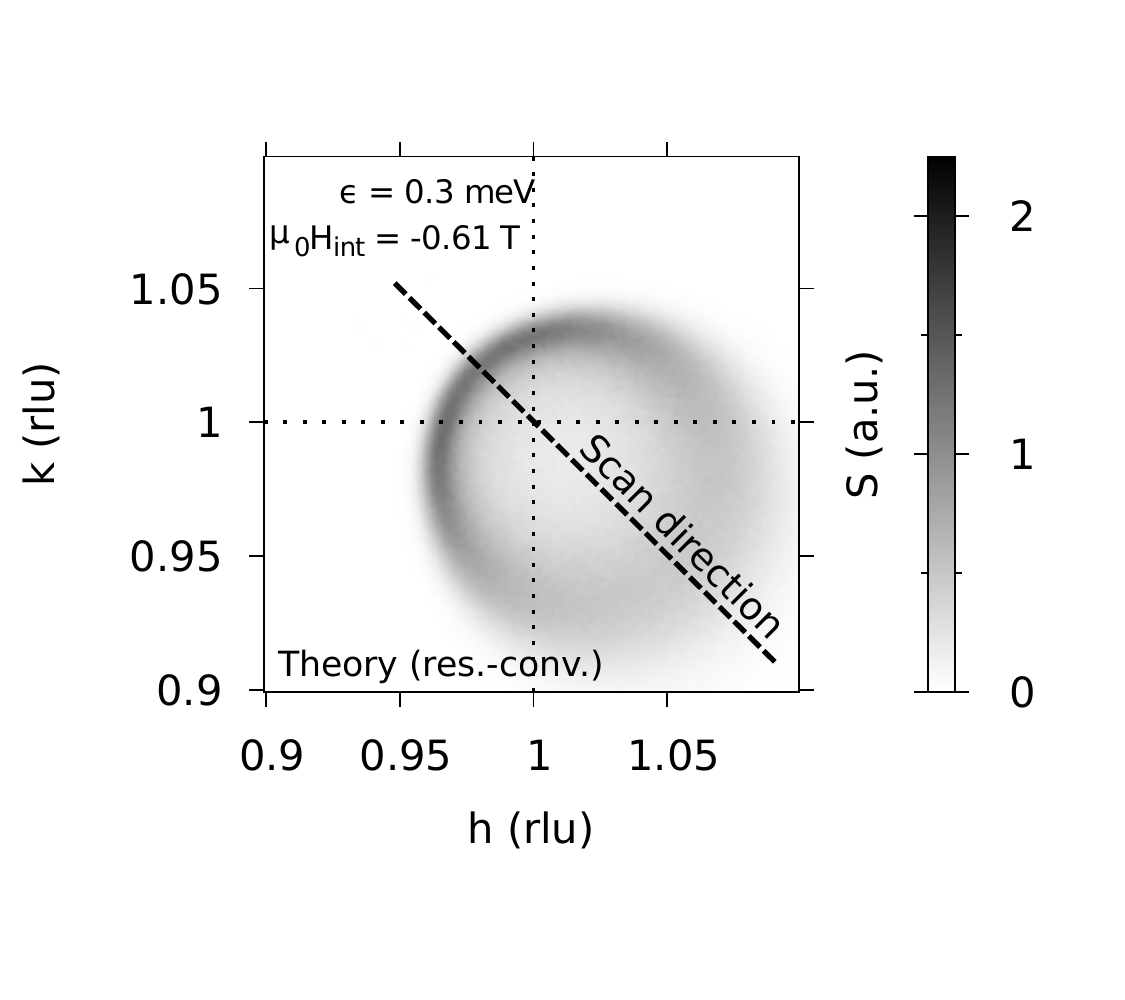} \hspace{10bp}
    \includegraphics[trim=4bp -15bp 25bp 10bp, clip, width=0.45\textwidth]{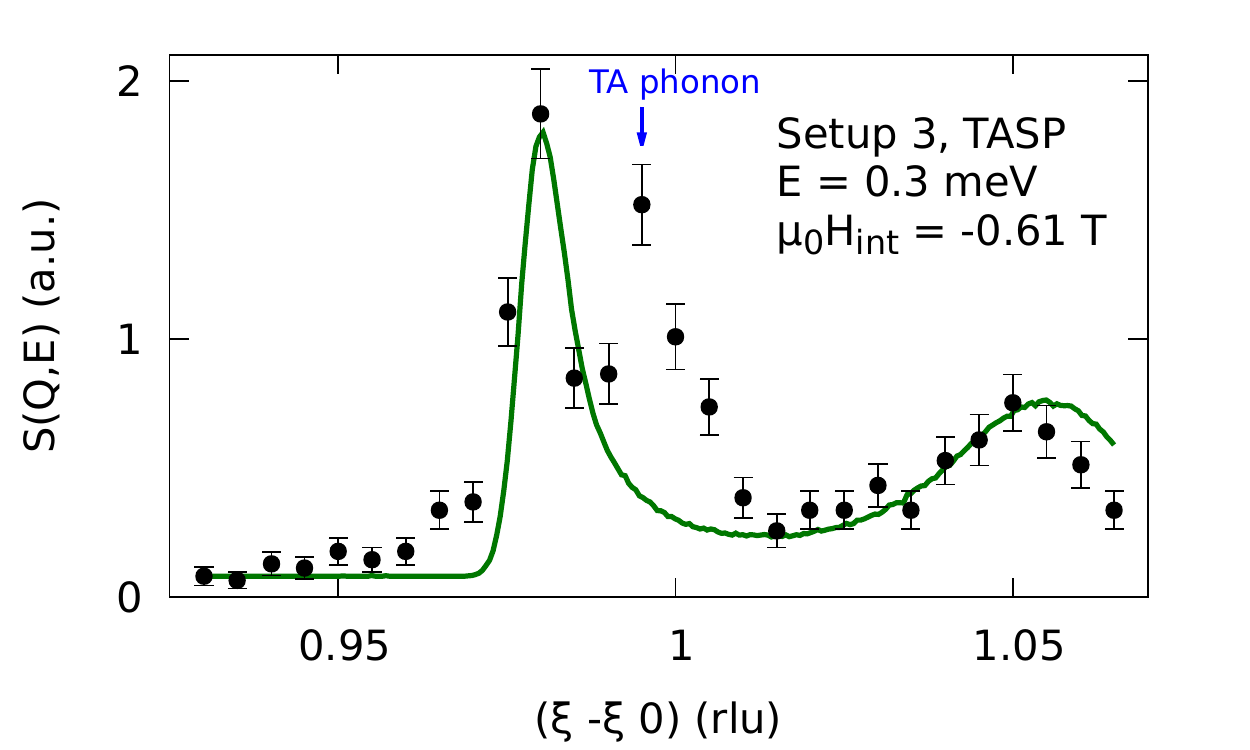}
    \caption{Field dependence of the non-reciprocal spin waves in the field-polarized phase for $\varepsilon({\bf q}) = 0.3\,\mathrm{meV}$ measured with $\bf{H} \perp {\bf{G}}$ (Setup 3). The left panel shows the convolution integral of the theoretical dynamical structure factor and the instrumental resolution function (blurred circle). The experimental results are depicted in the right panel. Here, two magnon modes and a transverse-acoustic (TA) phonon are visible. Only the magnon dispersion is modeled by the theory. The apparent broadening of one of the magnons is an instrumental resolution effect.}
    \label{setup3_E03}
\end{figure*}

Figs.~\ref{setup1_175} -- \ref{setup1_280} show constant-${\bf q}$ scans as measured using setup 1 (see Fig. 1 {\bf a} in the main text) for reduced momentum transfers $1.75\, k_h \leq |{\bf q}_\perp| \leq 2.8\, k_h$.
Figs.~\ref{setup2_200} and \ref{setup2_225} depict constant-${\bf q}$ scans as measured using setup 2 (Fig. 1 {\bf b} in the main text) for $|{\bf q}_\perp| = 2\, k_h$ and $|{\bf q}_\perp| = 2.25\, k_h$, respectively. In addition,
Figs.~\ref{setup3_220} and \ref{setup3_275} show constant-${\bf q}$ scans collected using setup 3 (Fig. 1 {\bf c} in the main text) for $|{\bf q}_\parallel| = 2.2\, k_h$ and $|{\bf q}_\parallel| = 2.75\, k_h$, respectively.
Fig.~\ref{setup3_E03} shows a constant-energy scan for an energy transfer of $\varepsilon = 0.3\,\mathrm{meV}$ performed using setup 3, illustrating the asymmetry of the magnon spectrum. Here, a transverse-acoustic (TA) phonon branch (cf. Ref.~\onlinecite{Lamago10}) was also observed. It was excluded for data analysis of the magnons.

In all figures, the experimental data are given as black and orange points, with black signifying the helimagnetic/conical phase and orange the field-polarized phase. The convolution integrals of the theoretical dynamical structure factor and the full four-dimensional instrumental resolution function \cite{Eckold2014} are given as green lines. Resolution-corrected contributions of individual helimagnon bands are indicated as blue lines. Note that the lines can appear jittery due to the Monte-Carlo approach towards solving the convolution integral. Details on the calculations and data treatment software are provided in Refs. \onlinecite{Takin2016, Takin2017, PhDWeber}.

In the constant-q scans for fields $\mu_0H_{int} = 87$ mT and $261$ mT in setup 3 (Fig.~\ref{setup3_220}) the sharp and intense signal at small energy transfer $E \simeq 0.025\,\mathrm{meV}$ was identified as a Bragg tail caused by the close proximity of the reduced momentum transfer $|{\bf q}_\parallel| = 2.2\,k_h$ to one of the magnetic satellite reflections. Due to its clear separation from the magnon peaks, the Bragg tail caused no complications during the course of the data analysis. For all other scans at larger momentum transfer $|{\bf q}_\parallel| > 2.2\,k_h$ no Bragg tail was observed.
Additional intensity was measured in some configurations for magnon energy loss of the (1') mode than is predicted by theory (see, e.g., the lower-right panel of Fig.~\ref{setup3_220}). This effect is no spurious Bragg tail as it is visible in the non-focusing direction of the spectrometer and warrants a further investigation in a future study.
In Fig.~\ref{setup1_225} a spurious signal caused by nuclear scattering from the sample environment is visible. These non-magnetic spurions posed no further problems for data analysis as they could be easily separated from the magnetic signals of interest.

\section{Theoretical analysis}

For the evaluation of the dynamical susceptibility $\chi''(\varepsilon, {\bf q})$ that enters the
scattering cross section of Eq.~(2) of the main text, we employed the linear spinwave approximation
of the theory presented in the context of Eq.~(1) of the main text. Details of the calculations in the
helimagnetically ordered phase were presented in the supplement of Ref.~\onlinecite{Schwarze15} as well
as in Ref.~\onlinecite{Max15}, see also the review of Ref.~\onlinecite{Garst2017}.

The magnon spectrum in the field-polarized phase, $H>H_{c2}$, can be obtained in closed-form and reads for a
left-handed magnetic system \cite{Garst2017}
\begin{align}
\varepsilon({\bf q})
= - 2 \mathcal{D} k_{h0} q_\parallel + \left(\left(\mathcal{D} \left(q^2 + \mathcal{A} \frac{q^4}{k_h}\right) + g \mu_B \mu_0 H_{\rm int}\right)\right.
\nonumber\\
\left.\left(\mathcal{D} \left(q^2 + \mathcal{A} \frac{q^4}{k_h}\right) + g \mu_B \mu_0 H_{\rm int}
+
\frac{g \mu_B \mu_0 m\, q_\perp^2}{q^2}
\right)\right)^{1/2}.
\end{align}
For the momentum we use the abbreviations $q = |{\bf q}|$ and $q_\perp = |{\bf q}_\perp|$, where ${\bf q}_\perp$
is perpendicular to ${\bf H}_{\rm int}$. Here, ${\bf H}_{\rm int}$ designates the internal magnetic
field. $q_\parallel = \hat H{\bf q}$ is the component of the wavevector $\bf q$ that is along the magnetic field.
Note that $q_\parallel$ can assume negative or positive values depending on the orientation of ${\bf H}_{\rm int}$
with respect to $\bf q$. The other parameters are defined in the main text.

The spectrum is non-reciprocal due to the first term that depends on the sign of $q_\parallel$.
Neglecting the higher-energy correction $\mathcal{A}$, the spectrum for a wavevector along the magnetic field
${\bf q} = q_\parallel \hat H$ reduces to the shifted parabola,
$\varepsilon(q_\parallel) \approx \mathcal{D} (q_\parallel - k_{h0})^2 + g \mu_B \mu_0 H_{\rm int} - \mathcal{D} k_{h0}^2$
that becomes gapless at the critical field $g \mu_B \mu_0 H^{\rm int}_{c2} =   \mathcal{D} k_{h0}^2$.
A small but finite $\mathcal{A}$ slightly renormalizes the parameters.

\begin{figure*}[!htb]
    \includegraphics[width=1\textwidth]{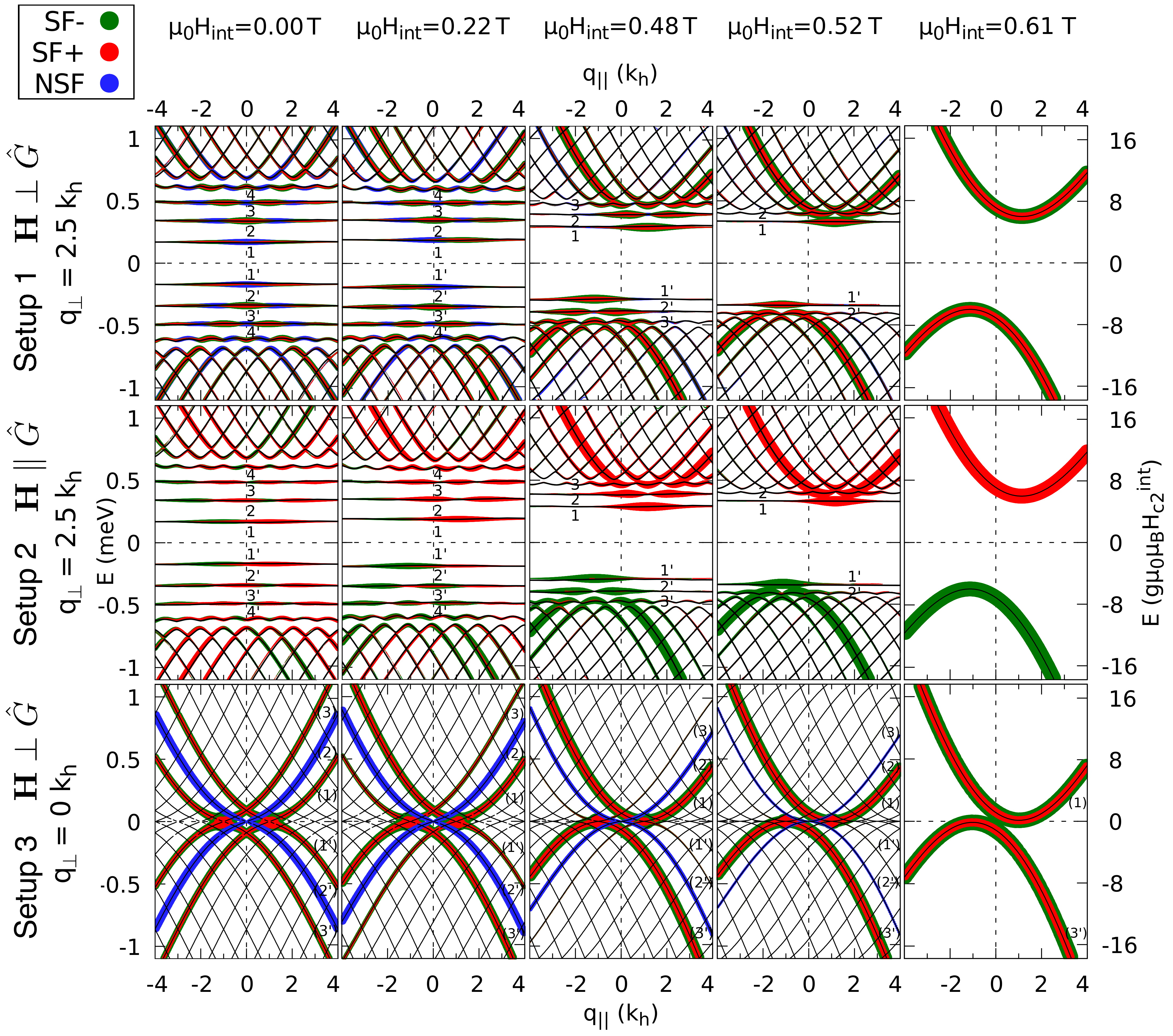}
    \caption{The field dependence of the three setups as predicted by helimagnon theory. It amends the magnon spectra shown in the main paper in Fig. 1.
    \label{theo}}
\end{figure*}

The theoretical field-dependent spectra for all employed setups are depicted in Fig. \ref{theo}.
The figure amends Fig. 1 of the main text with additional plots at several more field magnitudes.

A technical note on the implementation: While the full dispersion branches for $E>0$ and $E<0$ are directly obtained
in closed form for the field polarized phase, the model for the helimagnon phase as given in
\onlinecite{Schwarze15, Max15} originally only calculated the dispersion branches for $E > 0$. We modified it to
calculate the branches with respect to $E < 0$ by formally turning the field direction by 180 degrees in the code.

%
%

%
%

%
%

%
%

%

\section*{Author contributions}
T.W. planned and conducted the experiments. J.W. and M.G. created the theory. G.S.T. and R.G. were responsible for the TASP and MIRA instruments, respectively. A.B. and C.P. grew and characterized the sample. T.W. performed the data analysis. M.G. wrote the theoretical part of the manuscript, T.W. wrote the experimental part and created the figures \cite{PhDWeber}. P.B. contributed to the text. M.K. was involved in the initial interpretation of the results. P.B. supervised the entire work.